\documentclass[aps,prl,reprint,preprintnumbers,amsmath,amssymb,amsfonts,superscriptaddress]{revtex4-1}
\usepackage[latin1]{inputenc}
\usepackage[english]{babel}
\usepackage{graphicx}
\usepackage{xcolor}
\usepackage[%
  colorlinks=true,
  urlcolor=blue,
  linkcolor=red,
  citecolor=blue
]{hyperref}

\begin{document}
\title{An optical investigation of the strong spin-orbital coupled magnetic semimetal YbMnBi$_2$}

\author{Dipanjan Chaudhuri}
\affiliation{Department of Physics and Astronomy, The Johns Hopkins University, Baltimore, MD 21218, USA}

\author{Bing Cheng}
\affiliation{Department of Physics and Astronomy, The Johns Hopkins University, Baltimore, MD 21218, USA}

\author{Alexander Yaresko}
\affiliation{Max-Planck-Institute for Solid State Research, Heisenbergstrasse 1, 70569 Stuttgart, Germany}

\author{Quinn D. Gibson}
\affiliation{Department of Chemistry, Princeton University, Princeton, NJ 08544, USA}

\author{R. J. Cava}
\affiliation{Department of Chemistry, Princeton University, Princeton, NJ 08544, USA}

\author{N. P. Armitage}
\affiliation{Department of Physics and Astronomy, The Johns Hopkins University, Baltimore, MD 21218, USA}

\begin{abstract}
Strong spin-orbit coupling (SOC) can result in ground states with non-trivial topological properties. The situation is even richer in magnetic systems where the magnetic ordering can potentially have strong influence over the electronic band structure. The class of AMnBi$_2$ (A = Sr, Ca) compounds are important in this context as they are known to host massive Dirac fermions with strongly anisotropic dispersion, which is believed to be due to the interplay between strong SOC and magnetic degrees of freedom. We report the optical conductivity of YbMnBi$_2$, a newly discovered member of this family and a proposed Weyl semi-metal (WSM) candidate with broken time reversal symmetry. Together with density functional theory (DFT) band structure calculations, we show that the complex conductivity can be interpreted as the sum of an intra-band Drude response and inter-band transitions. We argue that the canting of the magnetic moments that has been proposed to be essential for the realization of the WSM in an otherwise antiferromagnetically ordered system is not necessary to explain the optical conductivity. We believe our data is explained qualitatively by the uncanted magnetic structure with a small offset of the chemical potential from strict stochiometry. We find no definitive evidence of a bulk Weyl nodes.  Instead we see signatures of a gapped Dirac dispersion, common in other members of AMnBi$_2$ family or compounds with similar 2D network of Bi atoms.   We speculate that the evidence for a WSM seen in ARPES arises through a surface magnetic phase.   Such an assumption reconciles all known experimental data.
\end{abstract}

\maketitle

\section{Introduction}
Correlated electron systems with strong SOC have been the subject of intensive research in recent years. The interplay of electronic correlations and SOC can result in emergent topological phases and has opened up a completely new direction in condensed matter physics. This interplay can be very different depending on the specifics of the electronic correlation. In weakly to moderately interacting electron systems, SOC can lead to non-trivial band topology as observed in conventional topological insulators \cite{moore2009topological}, Dirac and Weyl semi-metals \cite{Balents, turner2013beyond, armitage2017weyl}, axion insulators \cite{wan2012computational} and topological superconductors \cite{qi2011topological}. More recently, the effects of SOC on strongly correlated systems are being explored with the realization of new material systems with heavy 4d/5d transition metal compounds \cite{doi:10.1146/annurev-conmatphys-020911-125138}. The iridates deserve special mention in this category and have been instrumental in exploring much of this uncharted territory \cite{schaffer2015recent, rau2015spin}.

In addition to the emergence of topologically non-trivial ground states, the interplay between SOC and magnetic degrees of freedom themselves is also quite interesting. The family of AMnBi$_2$(A = Sr, Ca) compounds are particularly important in this context. Being structurally similar to iron based superconductors, they are referred to as manganese pnictides, which contain layers of Mn-Bi edge sharing tetrahedra and a Bi square net separated by a layer of A atoms \cite{guo2014coupling}. These compounds were expected from first principles DFT band calculations to host highly anisotropic Dirac dispersions with a finite gap at the Dirac point due to SOC \cite{lee2013anisotropic}. Such predictions were confirmed experimentally through magnetization, magneto-transport measurements \cite{park2011anisotropic, wang2012two} and later using angle resolved photoemission spectroscopy (ARPES) \cite{feng2013strong}. What makes these compounds particularly interesting is the antiferromagnetic (AFM) ordering of the spin magnetic moment on the Mn$^{2+}$ atoms which has a $3d^5$ electronic configuration. The magnitude of the ordered moment is around $4\mu_B$ with an ordering temperature $\sim$300K as observed in several experiments \cite{lee2013anisotropic, guo2014coupling}. Strong in-plane super exchange is responsible for the N\'eel-type AFM ordering observed in the $ab$ plane with a rather weak inter-layer magnetic coupling whose nature is dependent on the A atom \cite{guo2014coupling}. 

More recently, new members have been added to this family by substituting the rare earth elements with lanthanides such as Europium and Ytterbium. Whereas EuMnBi$_2$ appears to be very similar to its rare earth sister compounds \cite{PhysRevB.90.075109}, YbMnBi$_2$ appears to have richer behavior and has been proposed as a prospective candidate for type II Weyl semi-metal (WSM) \cite{borisenko2015time}. Unlike most WSMs discovered so far that are non-centrosymmetric systems (TaAs, NbAs, NbP, TaP, SrSi$_2$, etc) with broken inversion symmetry, YbMnBi$_2$ has been proposed to be a potential WSM with broken time reversal symmetry (TRS). A magnetic WSM has continued to be elusive to date, with the possible exception of potential WSM phases in pyrochlore iridates \cite{sushkov2015optical} and magnetic heuslers under magnetic field \cite{hirschberger2016chiral}. In contrast to the other members of the AMnBi$_2$ family, ARPES has shown a Fermi surface that appears to be continuous with the hole-like lenses touching the electron-like pockets at what has been interpreted to be the Weyl nodes. It should be noted, that the band structure calculations predict that a TRS breaking WSM state can only be realized in this compound if a canting of the magnetic moment ($\sim$10$^\circ$) from the $c$ axis, resulting in an effective in-plane ferromagnetic component along (110) direction, is assumed in addition to the established AFM ordering. Although such a band structure is in reasonable agreement with the ARPES results, the ad-hoc assumption of canted antiferromagnetism has not been supported by neutron scattering experiments \cite{FuhrmanPrivateCommunication, wang2016two}. Moreover, the magnetic space group found via neutron scattering that describes the AFM ordering in YbMnBi$_2$ has a symmetry breaking that does not allow canting of Mn magnetic moments away from the $c$ axis. From general considerations, the only way canting can occur is via an AFM transition that is not second order, or a lattice symmetry that is not tetragonal. Transport measurements reported in \cite{wang2016two} further supports a quasi two dimensional Dirac dispersion in this compound.

A number of interesting proposals have been made for the optical properties of WSMs \cite{hosur2013recent}. An ideal 3D Weyl state comprises at least a pair of non-degenerate band crossings in momentum space with the Fermi level at the nodal point. In 3D Dirac or Weyl systems, the joint density of states increases with frequency and is proportional to $\omega^2$ whereas the dipole optical matrix element goes as $\omega^{-1}$ as a result of which at low frequencies, the real optical conductivity is linear in $\omega$. In real systems however it is hard to achieve such fine tuning in the Fermi energy ($E_F$) and hence a cut-off is set by the Fermi level such that the real part of the conductivity is expected to be
\begin{equation}
\sigma_1(\omega) = \dfrac{N G_0 \omega}{24 v_F}\Theta(\omega-2E_F) \cdot \label{eq:lin}
\end{equation} 
where $N$ is the number of massless fermion species in the Brillouin zone, $G_0 = 2e^2/h$ is the quantum conductance and $v_F$ is the Fermi velocity. Experimental evidence of such linear optical conductivity with zero intercepts has been reported in Eu$_2$Ir$_2$O$_7$ \cite{sushkov2015optical} and TaAs \cite{xu2016optical} among others \cite{armitage2017weyl}. A more detailed theoretical analysis of the inter-band optical response of TRS breaking WSMs \cite{PhysRevB.93.085442} however predicts that the single linear optical conductivity should be broken into two regions of quasi-linear conductivity with different slopes by a peculiar kink. The kink is a manifestation of the van Hove singularity, appearing at a frequency that corresponds to the energy difference between the bands crossing to form the Weyl nodes at the extrema between the two nodes. Such a feature has been experimentally observed in the inversion symmetry broken WSM TaAs \cite{xu2016optical}. 

In this paper, we report the complex optical conductivity of YbMnBi$_2$ obtained from Kramers-Kronig (KK) constrained variational dielectric function (VDF) fitting of the reflectivity \cite{chaudhuri2016optical}. A comparison of the experimental optical spectra with that calculated from the DFT band structure shows that the low energy features in the spectrum can be attributed to specific inter-band transitions and are not a signature of the van Hove singularity as proposed elsewehere \cite{chinotti2016electrodynamic}. The optical spectra has also been compared to the calculated band structures with both canted and uncanted AFM ordering which opens the possibility of reconciling the ARPES experiments with the neutron scattering and transport results. We believe our optical data is explained qualitatively by the uncanted magnetic structure with a small offset of the chemical potential from strict stochiometry.  

\section{Experimental and Computational Details}

\begin{figure}[h!!]
\begin{flushleft}
\includegraphics[width =0.48\textwidth]{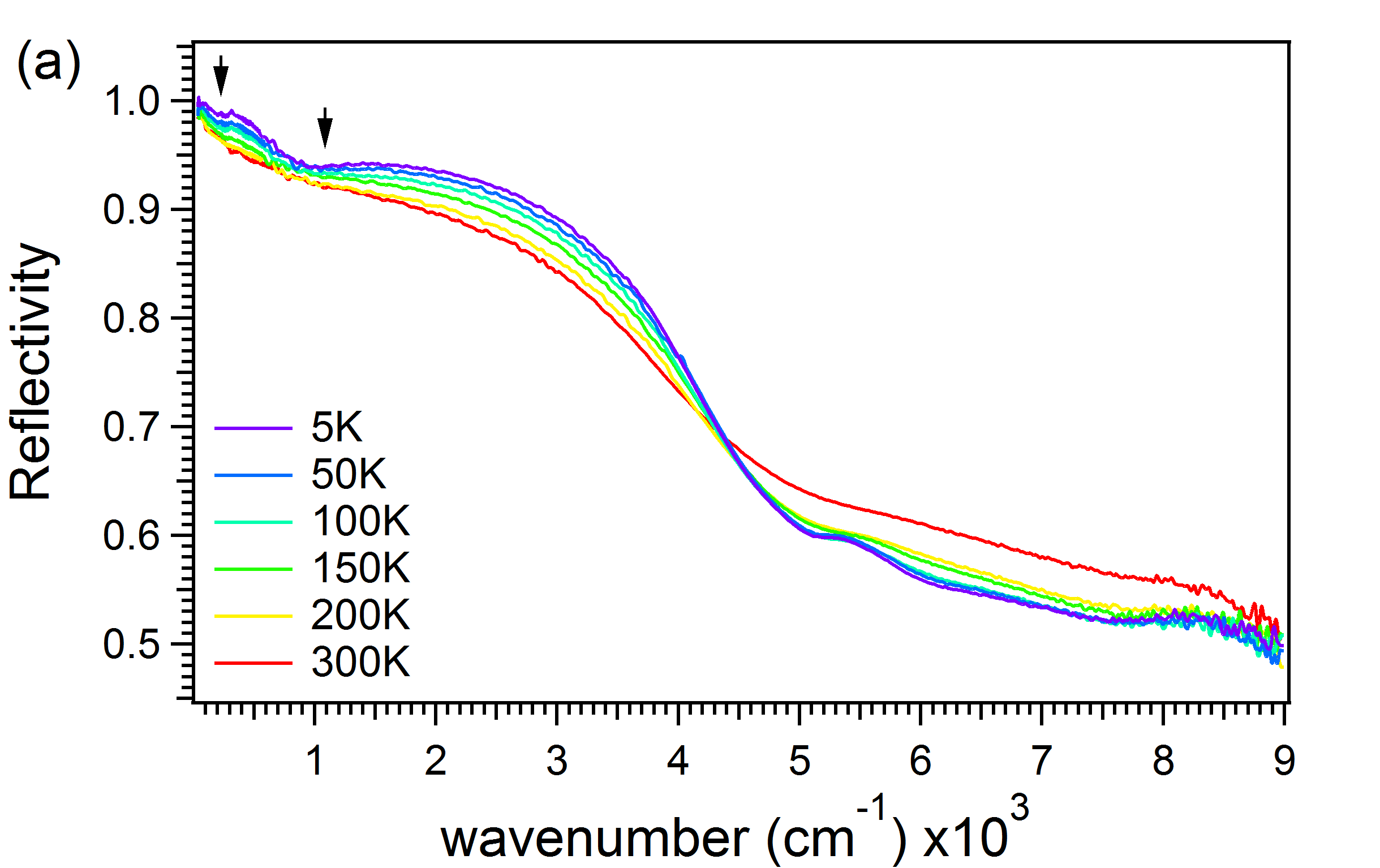}\\
\includegraphics[width=0.48\textwidth]{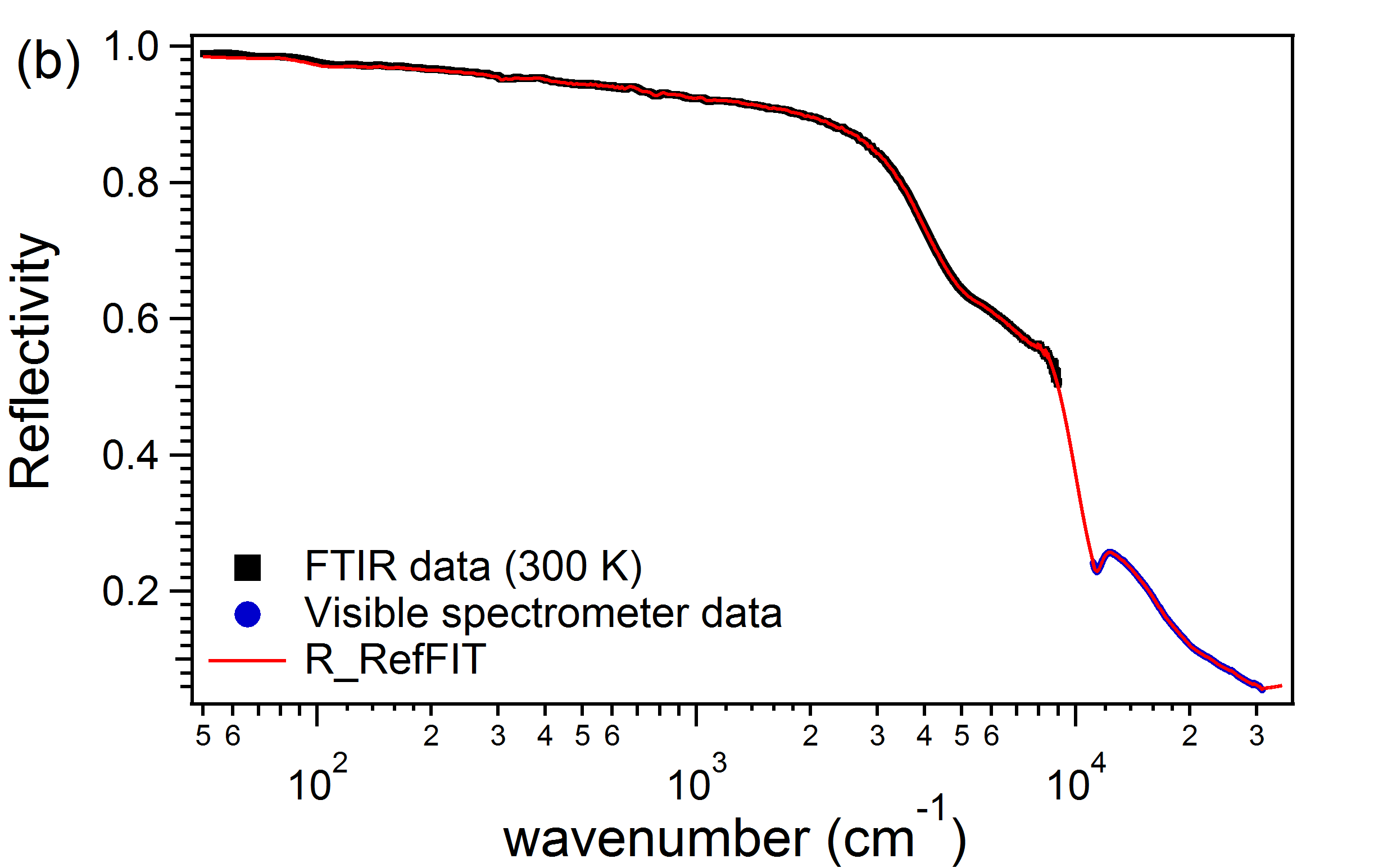}\\
\end{flushleft}
\includegraphics[width=0.53\textwidth]{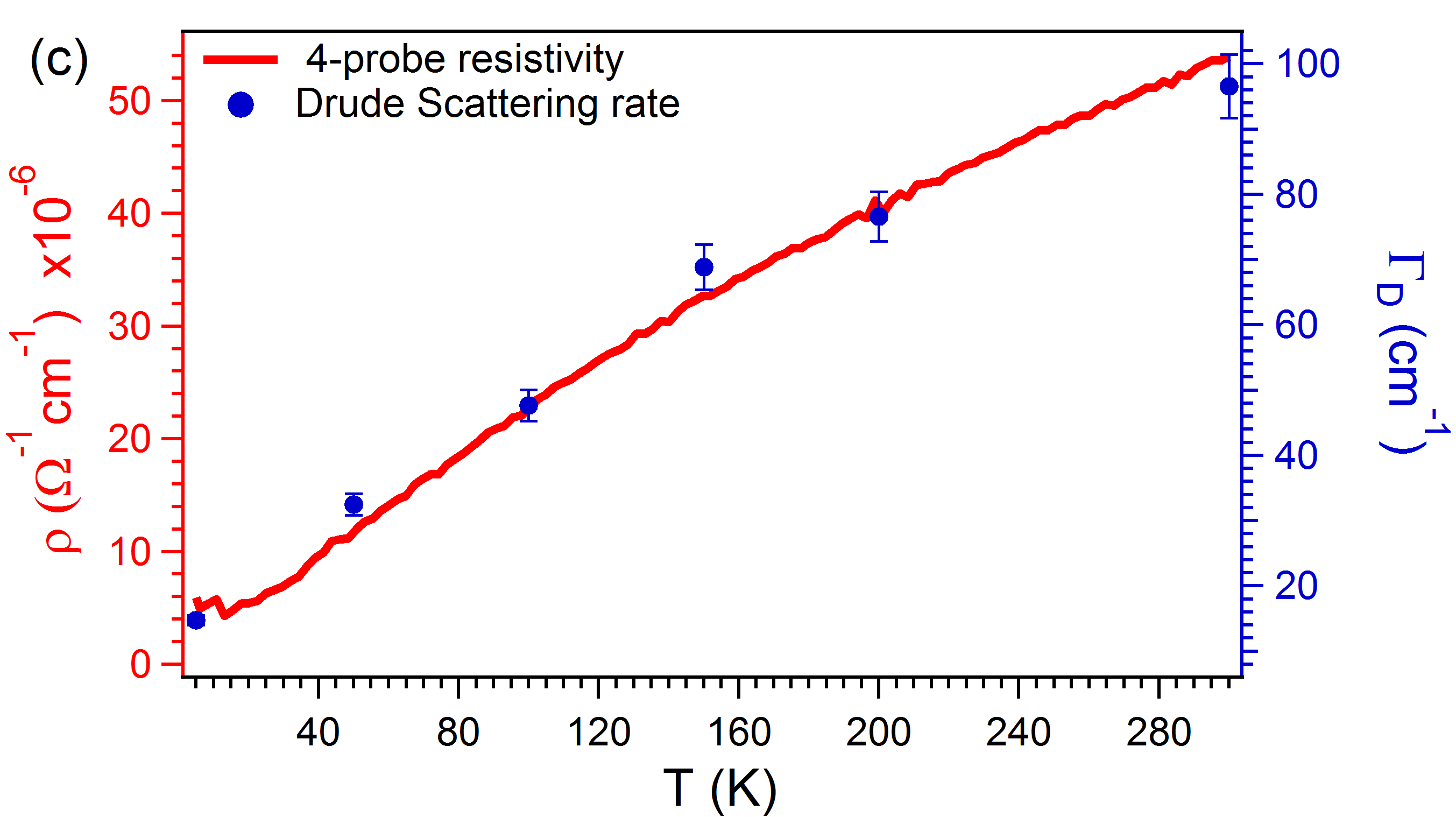}
\vspace{-10pt}
\caption{\label{fig:ref} (a) Reflectivity of YbMnBi$_2$ measured using FTIR spectroscopy. (b) Raw reflectivity data at 300 K along with reflectivity curve as obtained from VDF fitting. (c) Resistivity of the sample measured using a 4-probe technique along with the Drude scattering rates.}
\vspace{-10pt}
\end{figure}  

Optical properties were measured on a cleaved (001) surface ($\sim 3$ mm $\times$ 3 mm) of a high quality YbMnBi$_2$ crystal using Fourier transform infrared (FTIR) spectroscopy. For details of the sample preparation please refer to Ref.\cite{borisenko2015time}. The as-grown (001) surface of another slightly smaller crystal was also measured and had identical results. Fourier transform spectroscopy in known to offer excellent signal to noise and frequency resolution. Reflectivity of YbMnBi$_2$ single crystals were measured using a commercial FTIR spectrometer (Bruker, Vertex 80V, Source: Globar/Hg Arc Lamp, Detectors: DLaTGS/MCT/Bolometer) across far and mid infrared spectral ranges spanning from 50 cm$^{-1}$ to 8500 cm$^{-1}$ (i.e. 1.5-2500 THz) for several temperatures between 5 K and 300 K. The reflection spectra at each temperature were referenced to that of elemental gold, deposited on the sample in situ by thermal evaporation. To extend the measurement across a broader spectral range, the infrared spectra was supplemented by visible reflection spectrum between 11150 cm$^{-1}$ and 29000 cm$^{-1}$ measured at room temperature using commercial spectrophotometry (Hitachi, U-3010) and referenced to an aluminum mirror. Additionally, the DC resistivity of the sample has been measured using conventional 4-probe measurements down to 2K. 

Figure \ref{fig:ref}(a) shows the measured reflectivity in the infrared regime. The reflectivity tends to 1 at low frequencies and drops with a broad plasma edge-like feature above 4000 cm$^{-1}$ indicating metallic behavior. Besides the usual Drude response, two distinct features are observed in the reflectivity spectra that becomes particularly prominent at low temperatures. The lowest one is close to 200 cm$^{-1}$ where there is an abnormal increase in reflectivity and the second feature is a rather broad bump around 950 cm$^{-1}$.

Band structure calculations were performed using the linear muffin-tin orbital method \cite{And75} as implemented in the relativistic PY LMTO computer code. Some details of the implementation can be found in Ref.\ \onlinecite{book:AHY04}. Calculations were done assuming collinear AFM ordering of Mn moments which were aligned along the $c$ axis. Completely filled Yb$^{2+}$ 4$f^{14}$ states were treated as semi-core states. SOC was added to the LMTO Hamiltonian in the variational step. All theoretical results presented below were obtained within the local spin-density approximation (LSDA) with the Perdew-Wang parameterization \cite{PW92} for the exchange-correlation potential. Test calculations were also done using the PBE 
GGA potential. \cite{PBE96} They showed that the use of GGA slightly increases exchange splitting of Mn $d$ states but has only minor effect on Bi $p$ derived bands. Brillouin zone (BZ) integrations during the self-consistency loop were done on a 32$\times$32$\times$16 mesh using the improved tetrahedron method. \cite{BJA94}

Matrix elements for interband optical transitions were calculated in the dipole approximation. Then, the real part of the optical conductivity was calculated using the tetrahedron method and the imaginary part was obtained using the Kramers-Kronig relations.

We found that a very dense $k$ mesh should be used in order to achieve convergence of the conductivity below 0.3 eV. For instance, the in-plane conductivity spectrum calculated on the 32$\times$32$\times$16 mesh shows a broad peak centered at 0.18 eV. The peak shifts to $\sim$0.10 eV ($\sim$1000 cm$^{-1}$), becomes narrower, and its height doubles when a denser 80$\times$80$\times$48 mesh with almost 40000 symmetry inequivalent $k$ points in BZ is used. The use of even denser 128$\times$128$\times$48 ($\sim$10$^6$ $k$ points) mesh still leads to a small ($<$0.01 eV) shift of the peak position and 10\% increase of its height. However, since calculations on such a dense mesh becomes extremely time consuming we present in the next section conductivity spectra calculated
on the 80$\times$80$\times$48 mesh. It is worth noting, that above 0.5 eV convergence of the calculated conductivity spectrum is achieved already on the 32$\times$32$\times$16 mesh.

\section{Analysis and Results}
\begin{figure}[h!!]
\begin{center}
\includegraphics[width =0.48\textwidth]{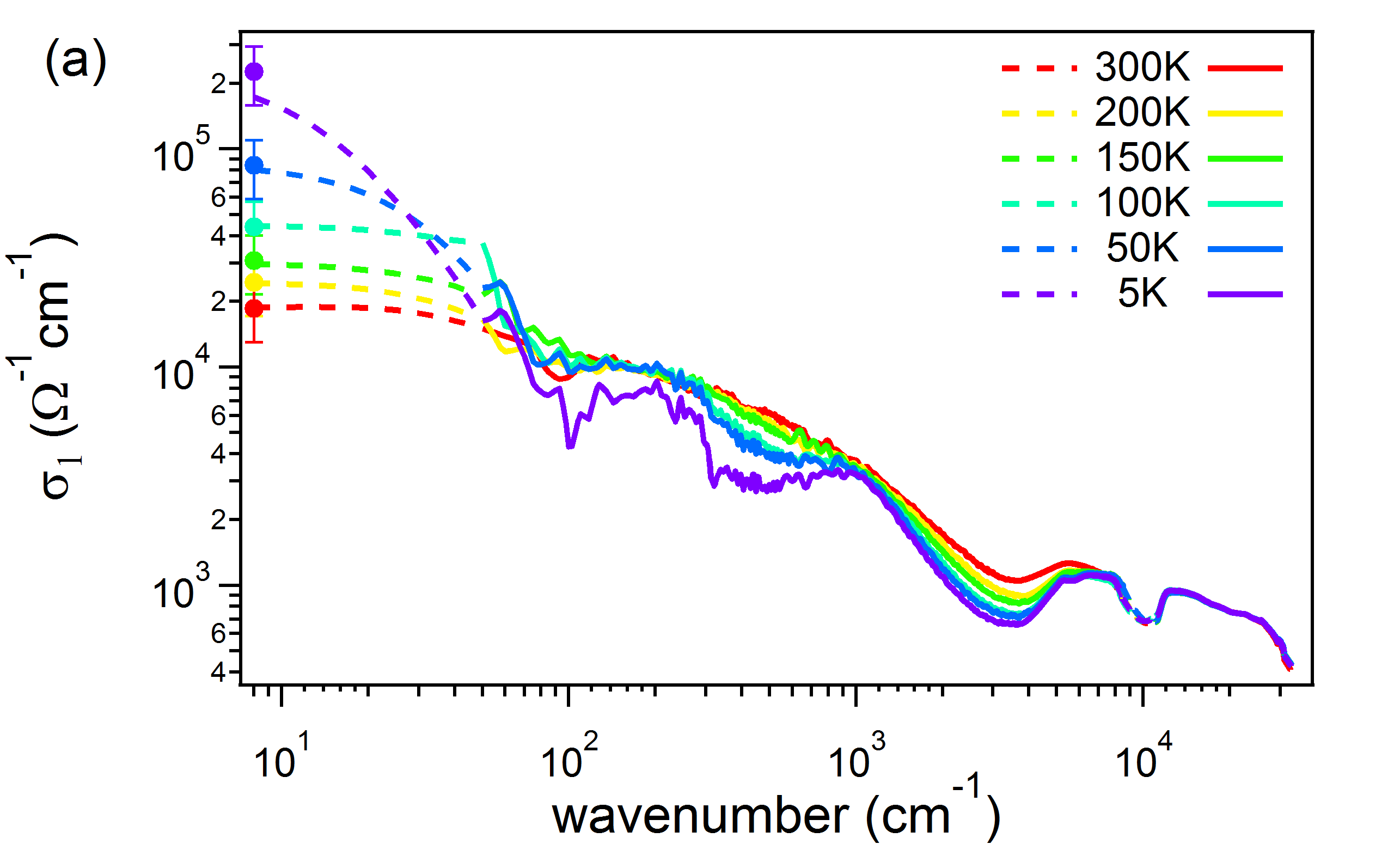}\\
\includegraphics[width=0.48\textwidth]{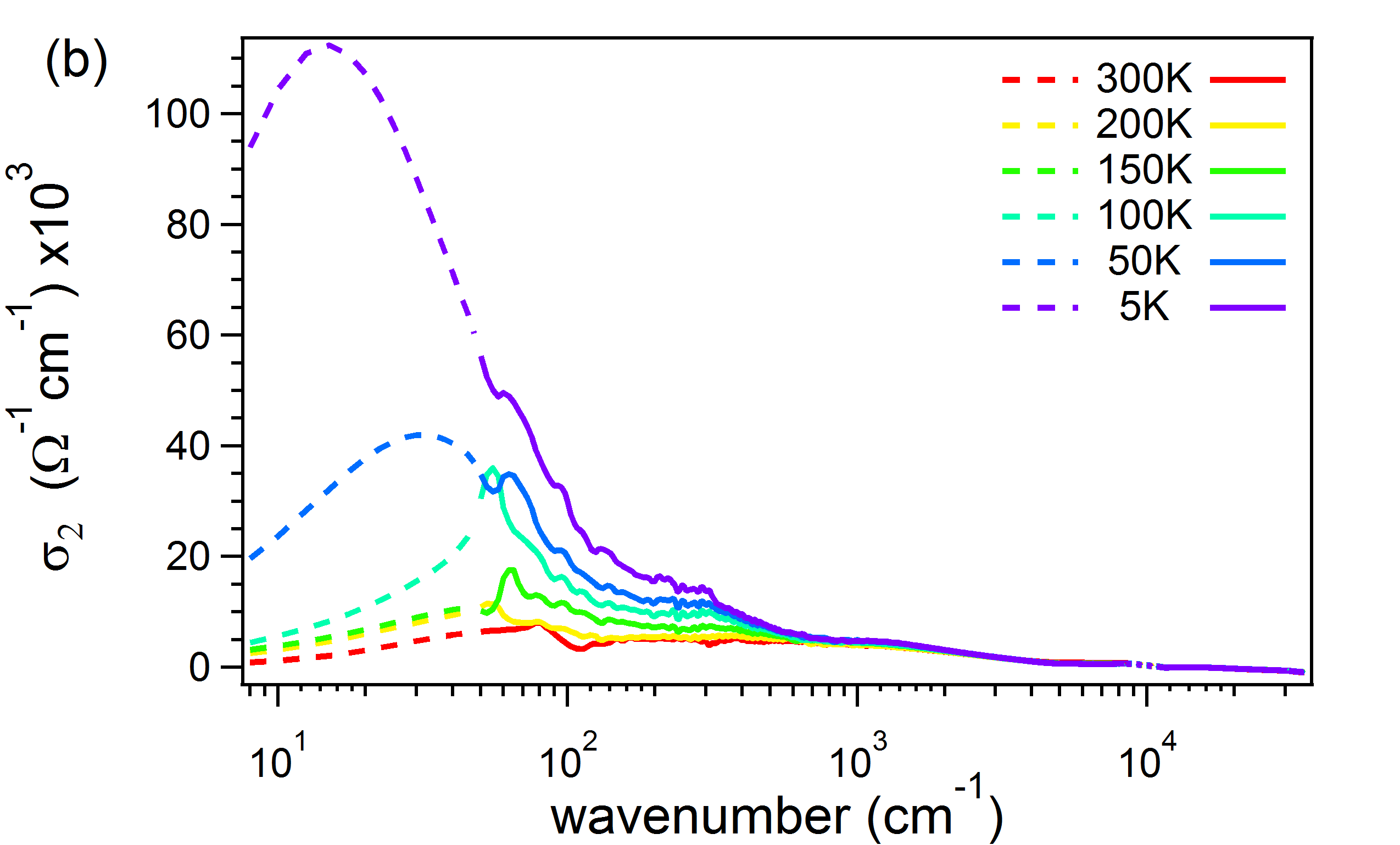}
\end{center}
\vspace{-10pt}
\caption{(a) Real and (b) imaginary part of conductivity. Solid and dashed lines represent the measured and extrapolated frequency regimes respectively. The markers at 8 cm$^{-1}$ in (a) represents the dc conductivity calculated from \ref{fig:ref}(c)}\label{fig:sigma}
\vspace{-10pt}
\end{figure}

The complex optical conductivity of the sample has been calculated from the reflectivity spectrum using KK constrained VDF fitting \cite{kuzmenko2005kramers}. This technique is an alternative to using conventional KK transformations to calculate the complex conductivity from a single measurement and is particularly useful in dealing with multiple reflection/transmission data from different disjoint frequency ranges. The method, implemented using the software RefFIT, involves an initial fitting of the reflectivity spectra to a standard Drude-Lorentz oscillator model with a few oscillators and then performing a KK constrained variational fitting with many oscillators of the difference spectra. For YbMnBi$_2$, the reflectivity spectra for each temperature has been fitted with one Drude and six Lorentz oscillators which has then been subjected to the VDF fitting analysis. Figure \ref{fig:ref}(b) shows the actual infrared and visible reflectivity data at 300K along with the output of the VDF fitting. In addition to reflectivity, the dc conductivity at each temperature as obtained from 4-probe resistivity measurements (see figure \ref{fig:ref}(c)) have also been used to constrain the VDF fitting subroutine. The real and imaginary parts of the complex conductivity thus obtained are shown in figure \ref{fig:sigma}(a) and \ref{fig:sigma}(b) respectively.

We model the low frequency response ($<$150 cm$^{-1}$) for all temperatures by a single Drude oscillator. Figure \ref{fig:ref}(c) shows the Drude scattering rate, $\Gamma_D$, as a function of temperature. Note that the temperature dependence of $\Gamma_D$ scales as resistivity within experimental error. At low temperatures, the resistivity of YbMnBi$_2$ has a strong temperature dependence showing the dominant role of inelastic scattering from either electron-electron or electron-phonon interactions. The similarity in temperature dependence of resistivity and $\Gamma_D$ thus points to the coherent nature of the quasiparticles within the Drude subsystem. The measured Drude plasma frequency at 5K is 1.74 eV.

The two most prominent features in the real part of the optical conductivity are the low energy peaks around 200 cm$^{-1}$ and 950 cm$^{-1}$. Such peaks in conductivity are ordinarily attributed to inter-band electronic transitions. To investigate this further, the Drude contribution (which mostly captures the intra-band optical conductivity) was subtracted. In general, strong electronic correlation can give low frequency intra-band contributions that would not generally be captured through a Drude term only. One possible way to isolate such effects is through the extended Drude model \cite{allen1977optical} analysis which can account for inelastic scattering due to electron-electron or electron-phonon interactions that can be frequency dependent even at low frequencies. However, the extended Drude model is only reliable much below the energy scales of inter-band transitions and thus it is not particularly useful in this context where the lowest possible inter-band transition could be as low as 200 cm$^{-1}$. We will therefore ignore such effects for now and assume that all the remaining high frequency optical conductivity is primarily from inter-band transitions and compare it to the inter-band optical conductivity calculated from band structure.

\begin{figure}[h!!]
\begin{flushright}
\includegraphics[width =0.45\textwidth]{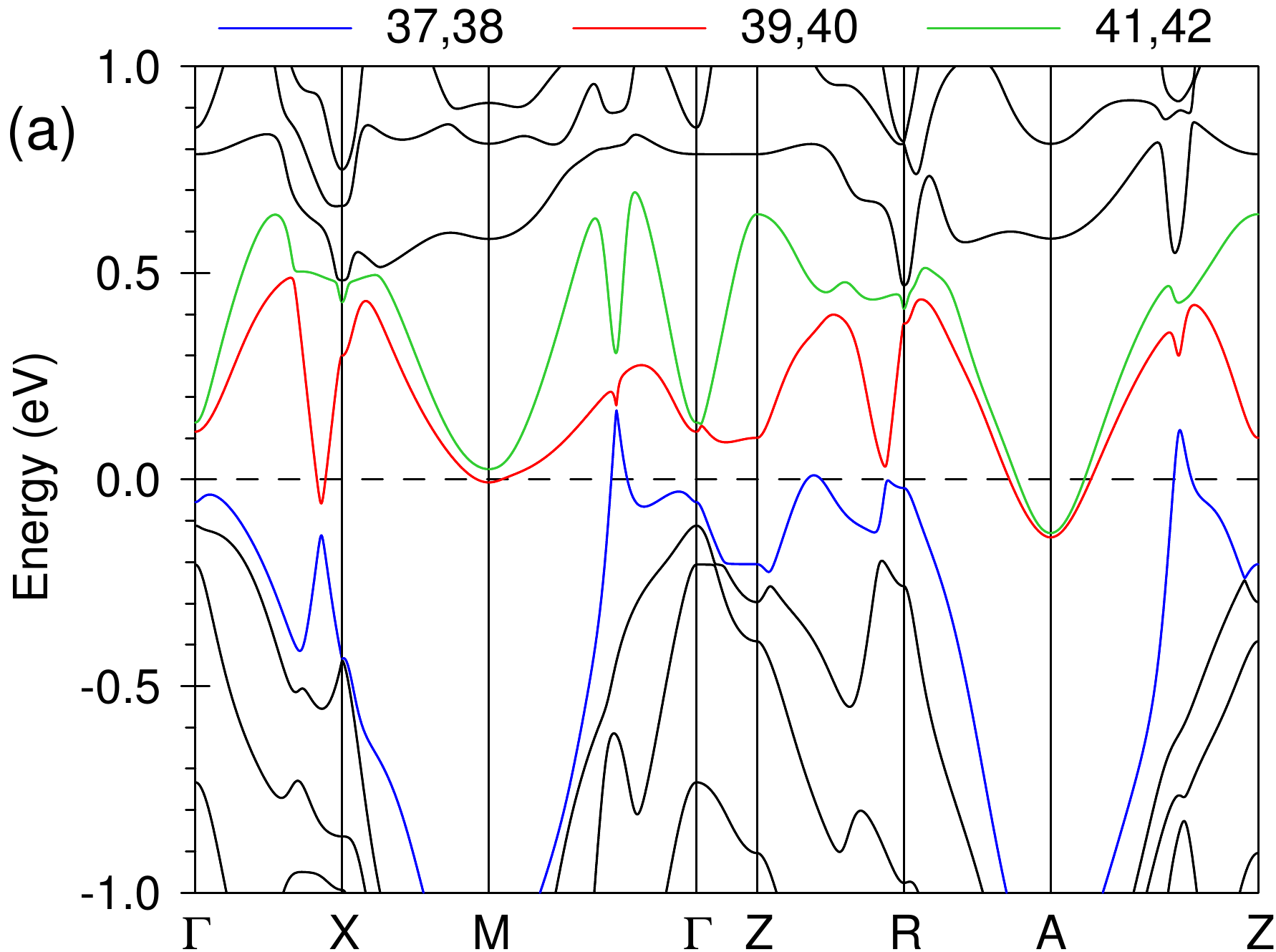}\\
\hspace*{1.5em}\includegraphics[width=0.48\textwidth]{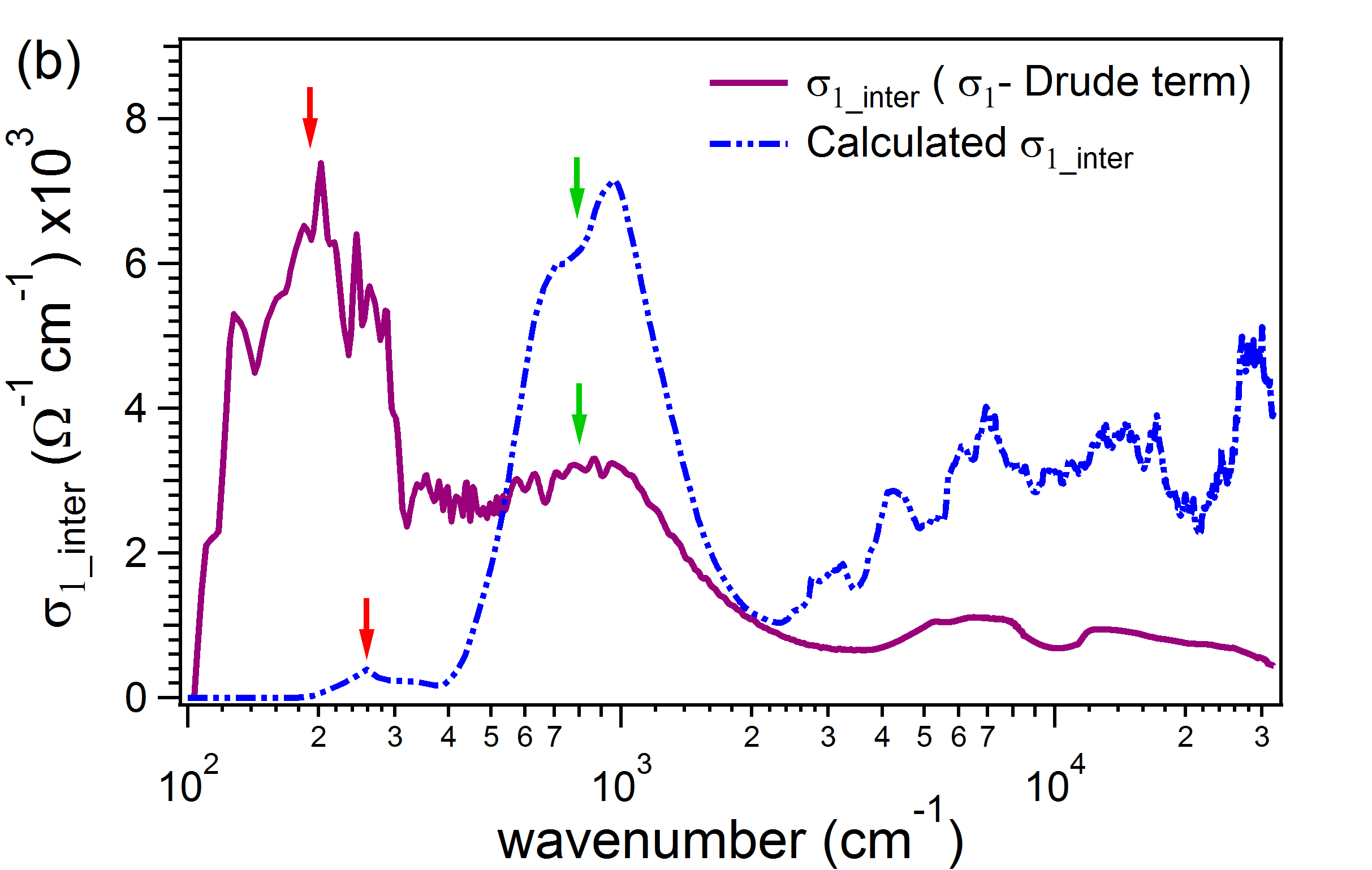}
\end{flushright}
\vspace{-10pt}
\caption{(a) Band structure calculated along symmetry lines in tetragonal BZ. Bands crossing the Fermi level, which are discussed below, are plotted with blue (37,38), red (39,40), and green (41,42) lines. and (b) comparison of the calculated optical conductivity with experiment}\label{fig:bs}
\vspace{-10pt}
\end{figure}

Calculated inter-band optical conductivity (see figure \ref{fig:bs}(b)) is in rough agreement with the experimental observations at low frequencies and successfully reproduces a peak at 950 cm$^{-1}$ (green arrows) although the predicted amplitude is much larger. However, as one can see, the calculated conductivity is significantly higher than the experimental one almost all throughout the frequency range. The origin and significance of this peak will be discussed further below. 

The more striking feature is perhaps the almost complete absence of a large peak close to 200 cm$^{-1}$ (red arrows) in the calculated optical conductivity spectra which might be indicative of some deeper inconsistencies. As the band structure in figure \ref{fig:bs}(a) does not include the canting of the Mn$^{2+}$ moments that is believed to be necessary for the magnetic WSM phase, a next obvious step is to include that in the DFT calculations. 

\begin{figure}[h!!]
\begin{flushright}
\includegraphics[width =0.45\textwidth]{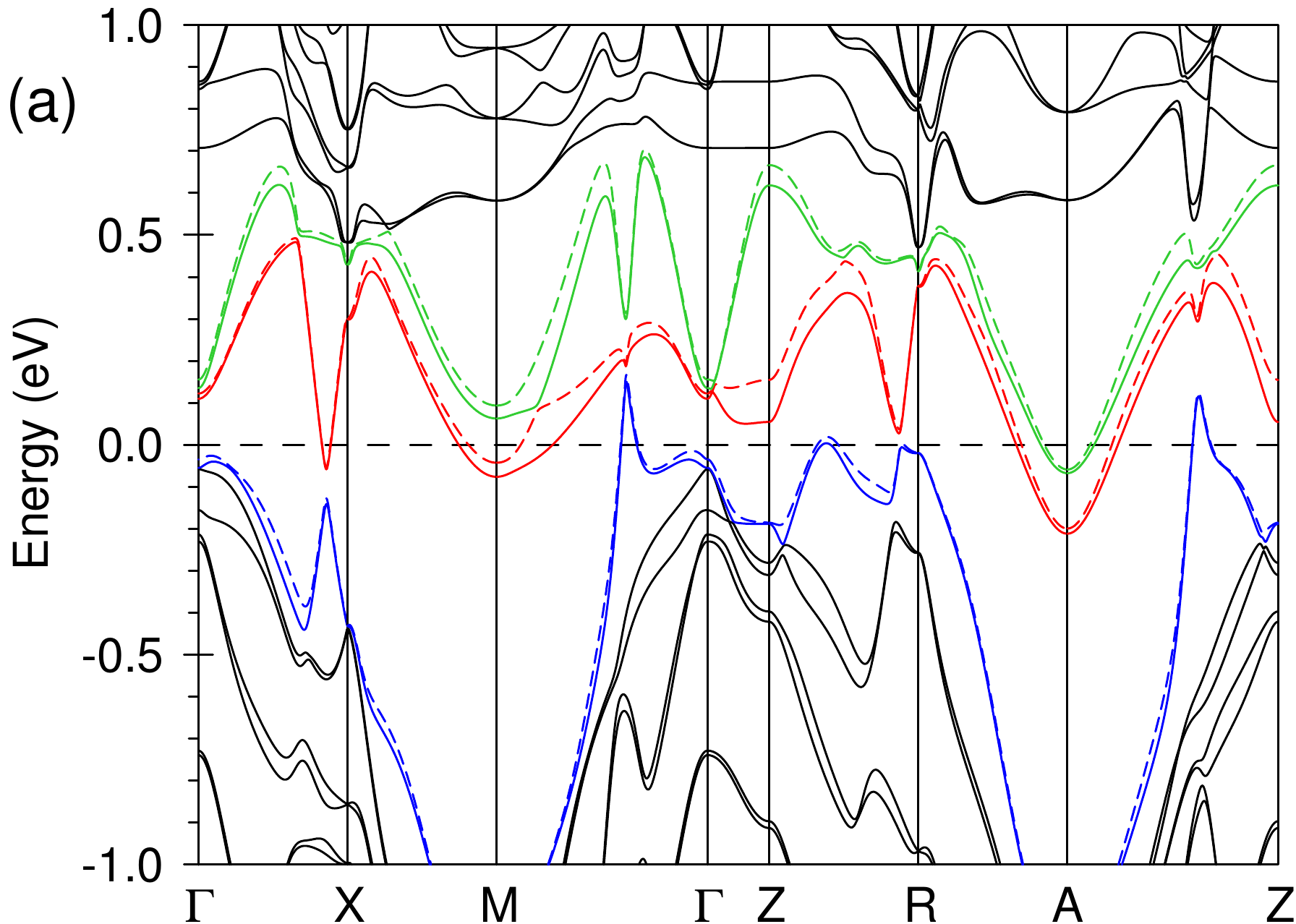}\\
\includegraphics[width =0.45\textwidth]{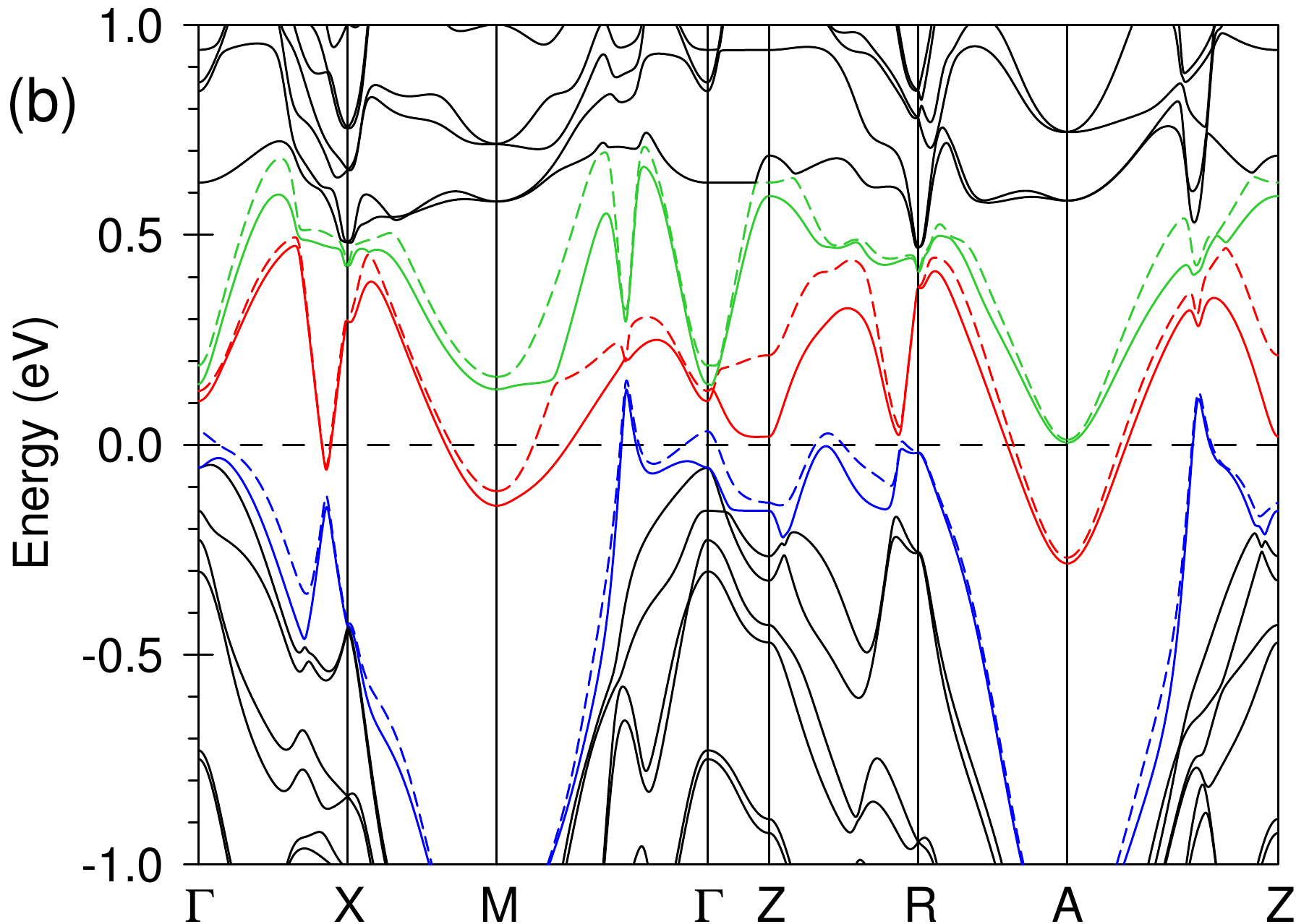}\\
~~~~~\includegraphics[width=0.48\textwidth]{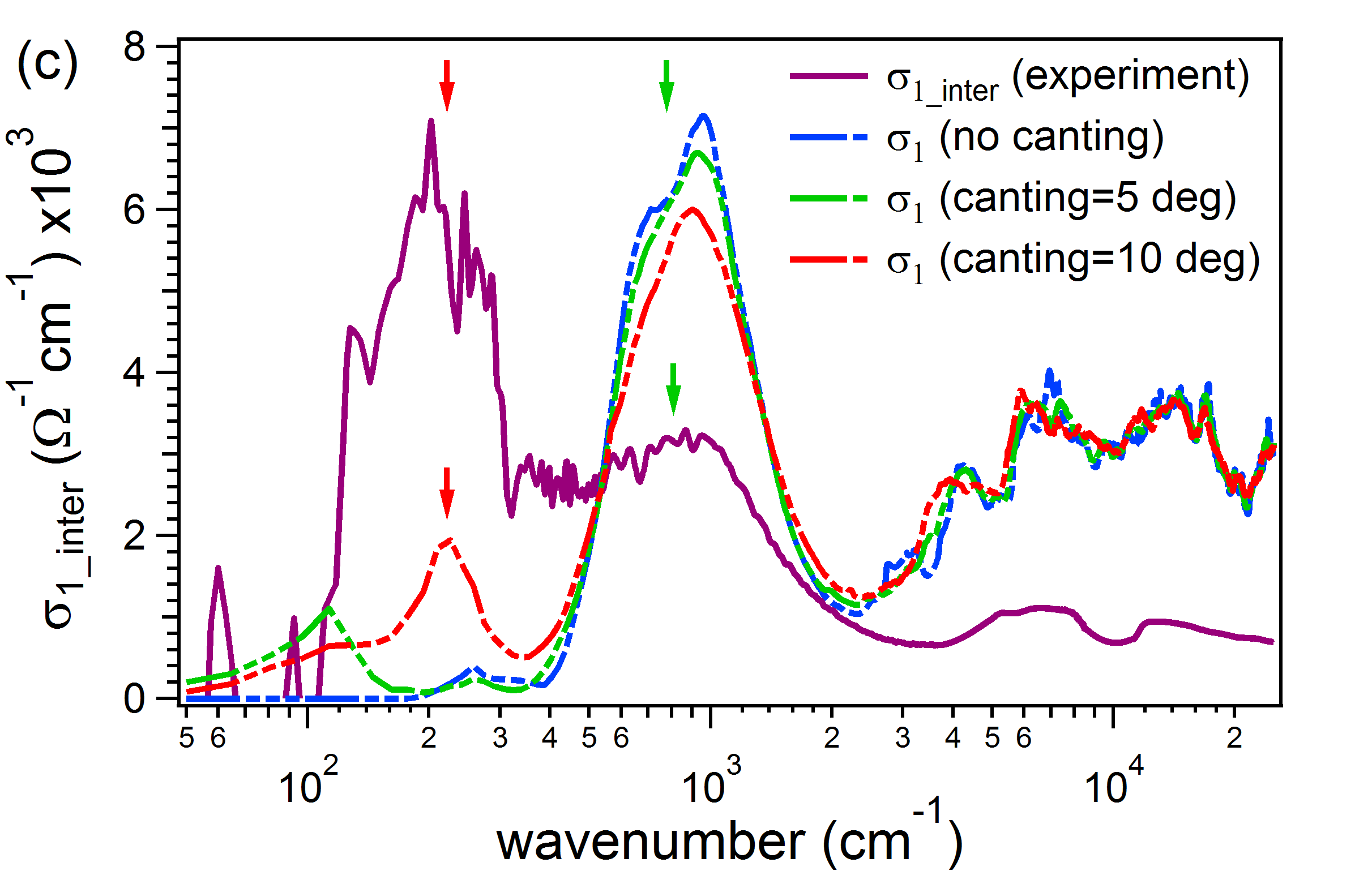}
\end{flushright}
\vspace{-10pt}
\caption{Band structure calculated along symmetry lines for a canting of (a) 5$^{\circ}$ and (b) 10$^{\circ}$ from the c-axis. The solid and dashed lines of the same colors represent the bands that were degenerate without canting. (c) Comparison of the calculated optical conductivity for different canting with experiment}
\label{fig:cant}
\vspace{-10pt}
\end{figure}

Figure \ref{fig:cant}(a) and \ref{fig:cant}(b) shows the band structure along (001) for a canting of 5$^{\circ}$ and 10$^{\circ}$ from the c-axis respectively. Two sets of Weyl points develop in the band structure for the canted system. The first set consists of two pair of Weyl nodes at momentum coordinates (0.193,0.193,0.12), (0.193,0.193,-0.09) and symmetry related points.  In figure \ref{fig:cant}(a) and \ref{fig:cant}(b), along the $k$-space cut along the M-$\Gamma$ line, the Weyl point can be seen to be
developing near 200 meV. The other four Weyl nodes are at (0.394, 0.045, 0.131) and symmetry related points. Details of these Weyl node structures have been discussed in reference \citep{borisenko2015time}. Note that for all points, there are certain non-idealities (both in distortions of the bands and their position from E$_F$) when compared to the simplest Weyl band structure proposed in \cite{PhysRevB.93.085442} which would make the observation of Weyl physics in the optical response challenging. 

The inter-band conductivity for these three levels of canting, as well as the experimental data is plotted in figure \ref{fig:cant}(c). Note that a peak slightly above 200 cm$^{-1}$ gradually develops as one includes the effect of canting in the band structure. Nevertheless, one must be cautious in claiming this to be evidence for the WSM phase as there are noticeable alterations in the band structure unrelated to WSM physics that might bring about the observed changes in optical conductivity. The principal modification to the band structure from canting is the lifting of two fold degeneracy of the bands close to the Fermi level.  This is the feature that gives rise to the  Weyl nodes in the calculated band structure. However, this also opens up the possibility of having transitions between these previously degenerate bands as not all of them are above or below the Fermi level. The signatures of such inter-band transitions are not expected to be directly related to the existence of the Weyl nodes 	as one can see that they are generally far from the Fermi level.

\begin{figure}[h!!]
\begin{flushright}
\includegraphics[width=0.45\textwidth]{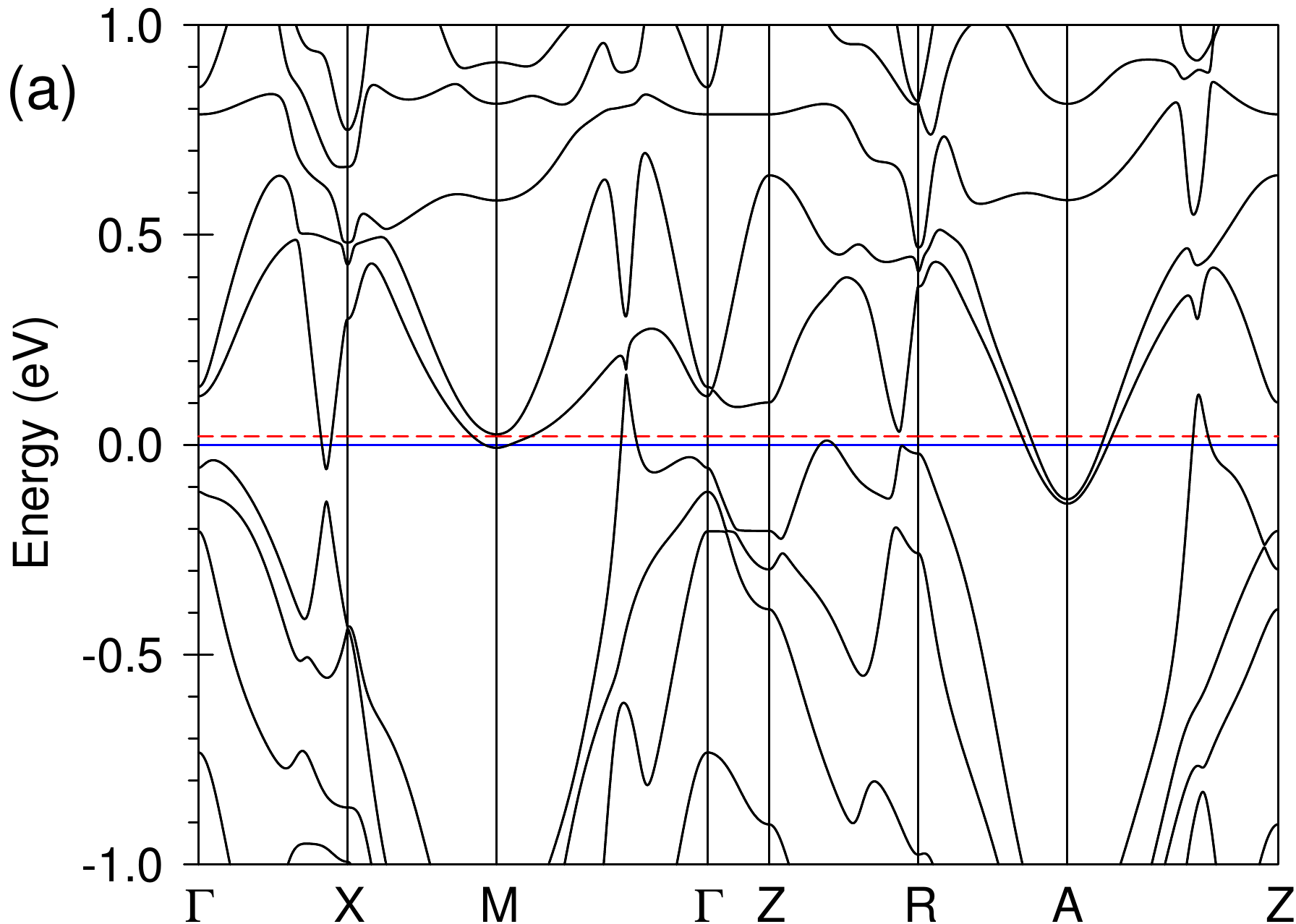}\\
\hspace*{1.5em}\includegraphics[width=0.48\textwidth]{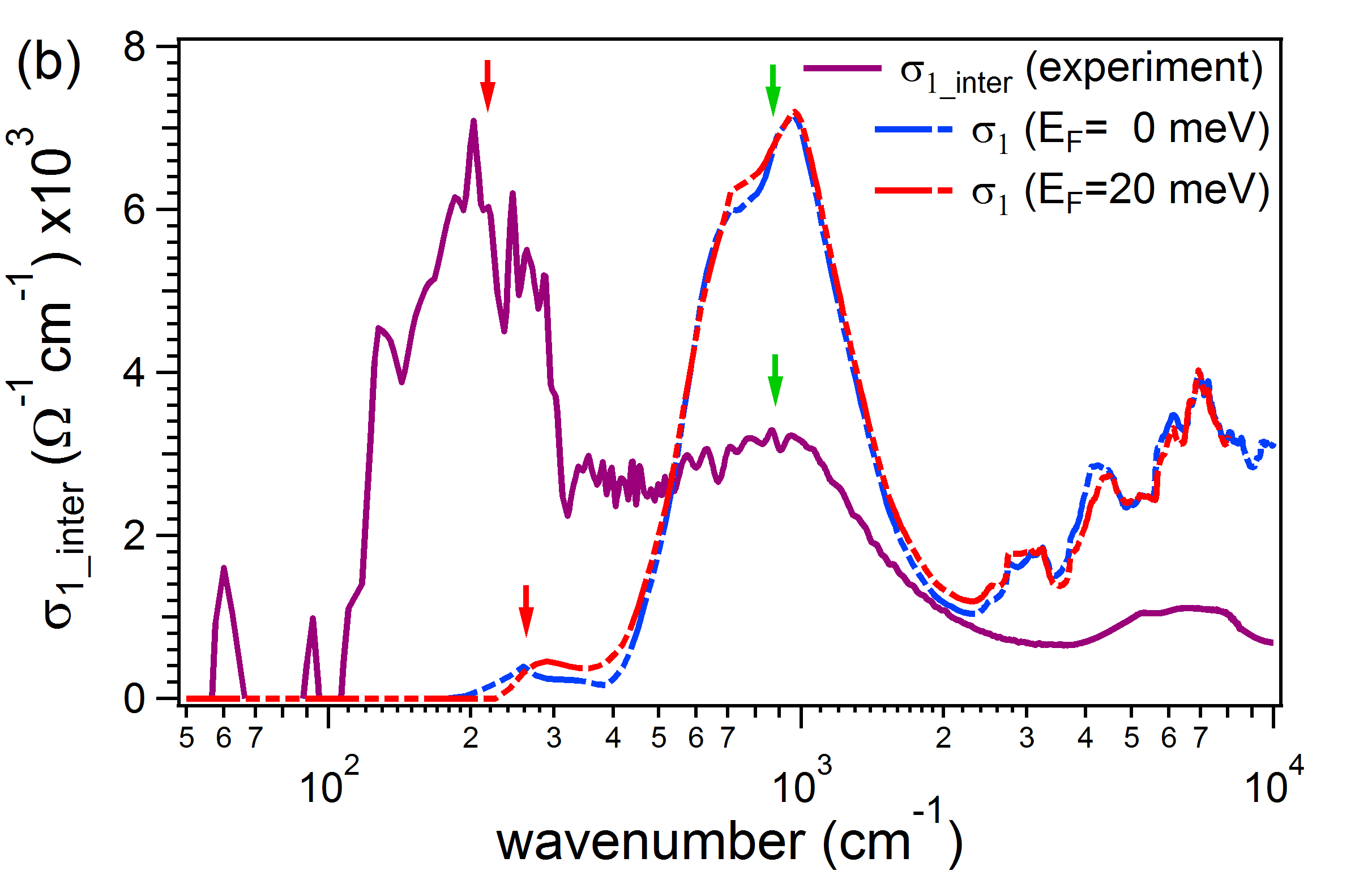}
\end{flushright}
\vspace{-10pt}
\caption{Band structure along (001) with Fermi level at 0 meV (blue solid line) and 20 meV above (red dashed line). (b) Calculated optical conductivity for the corresponding Fermi levels.}\label{fig:fermi}
\vspace{-10pt}
\end{figure}

If the important changes to the optical conductivity are indeed from the bands being rearranged around the Fermi level, somewhat similar changes may be observed by a shift of the chemical potential. Moreover, doping is rather common in semimetallic systems and thus exploring this possiblity is important. This was investigated by shifting the chemical potential above the calculated charge neutral Fermi level (see figure \ref{fig:fermi}(a)). An enhancement in the spectral weight under the calculated peak at 200 cm$^{-1}$ is seen upon shifting chemical potential by 20 meV above the charge neutral Fermi level (see figure \ref{fig:fermi}(b)). Thus it is possible that that the observed peak at 200 cm$^{-1}$ is simply indicative of a low energy inter-band transitions and not a signature of Weyl nodes as postulated elsewhere \cite{chinotti2016electrodynamic}.

\begin{figure}[h!!]
\begin{center}
\includegraphics[width=0.48\textwidth]{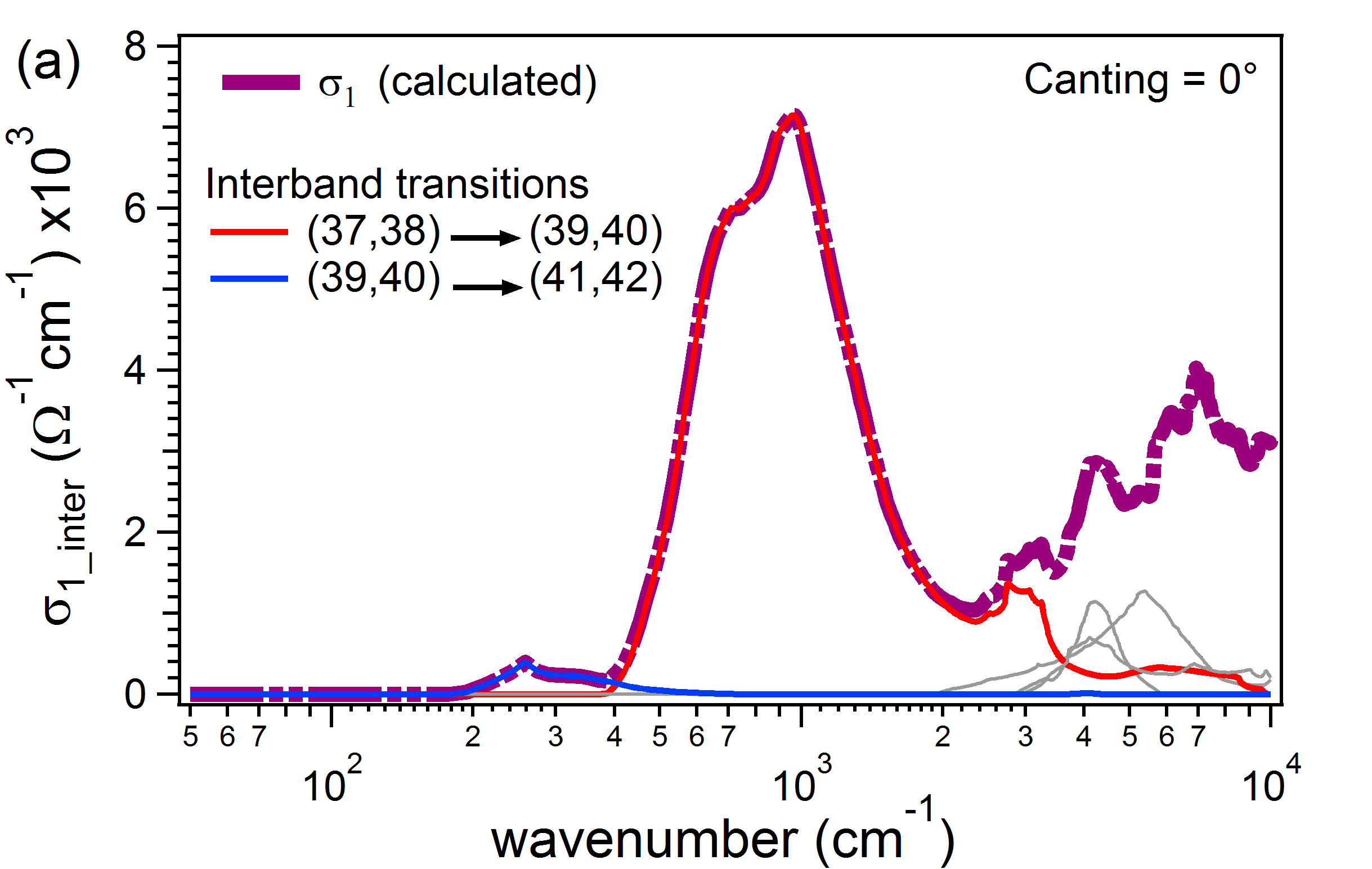}\\
\includegraphics[width=0.48\textwidth]{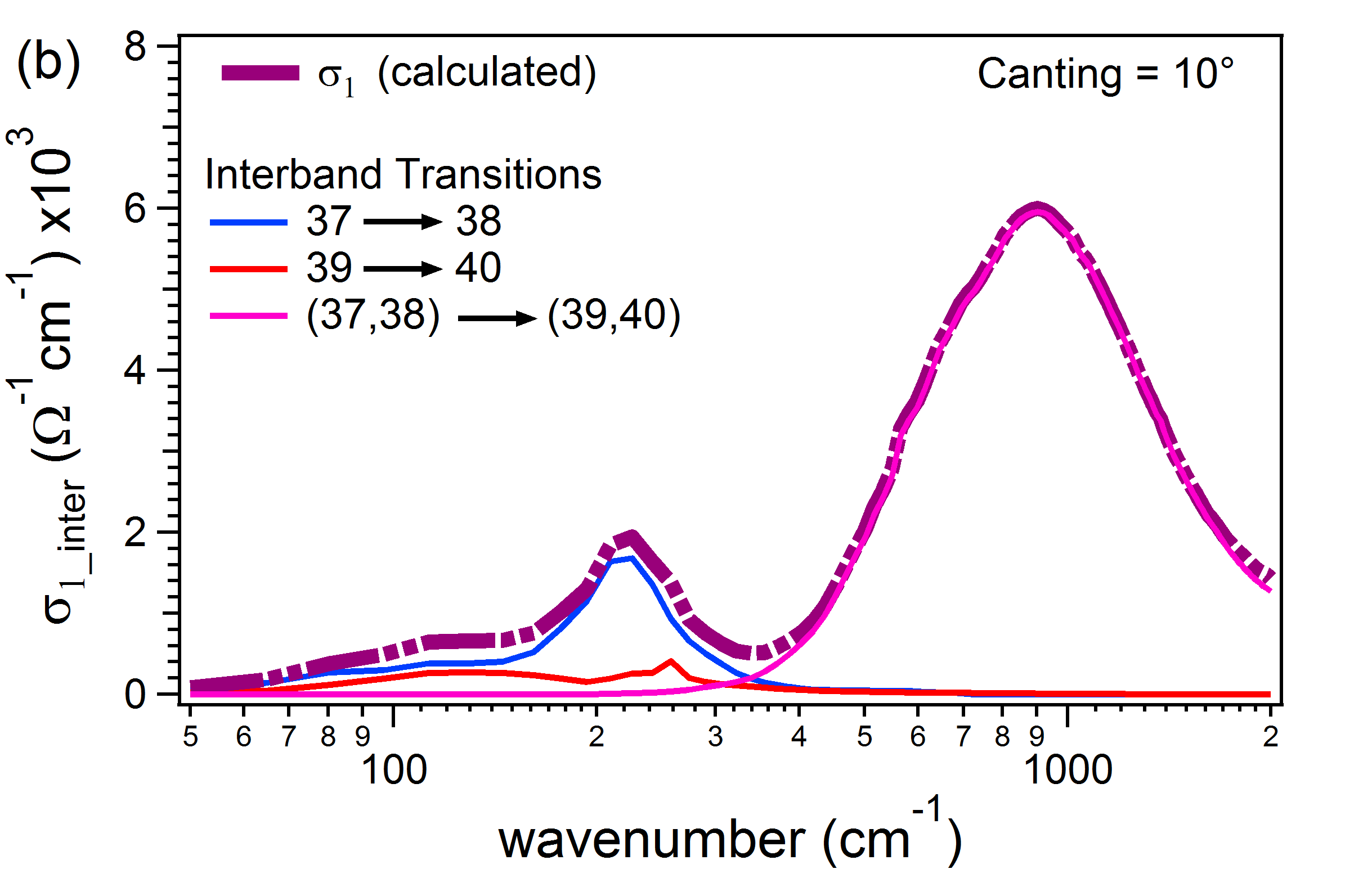}
\end{center}
\vspace{-10pt}
\caption{(a) Decomposition of the the total inter-band optical conductivity into individual band transitions. Transitions from bands (39, 40) to (41, 42) result in the peak at 200 cm$^{-1}$ whereas transitions from bands (37, 38) to (39, 40) is the origin of the peak at 950 cm$^{-1}$. (b) Similar decomposition assuming a 10$^{\circ}$ canted AFM order in the band structure.}\label{fig:dec}
\vspace{-10pt}
\end{figure}

To investigate the origin of these two peaks more closely, we decomposed the calculated conductivity into individual inter-band contributions. Figure \ref{fig:dec}(a) shows the result of that decomposition. These calculations provide more definitive proof to support that the peak in conductivity at 200 cm$^{-1}$ is indeed an inter-band transition from the set of degenerate bands that is mostly above the Fermi level (bands 39, 40) to the immediate higher one (bands 41, 42). This is why when the Fermi level is shifted to higher energies, the occupation in the lower degenerate set of bands (39, 40) and consequently the transition probability increases. Hence we observe the enhancement of the amplitude of the peak in conductivity upon shifting the Fermi level and is a probable cause of the experimentally observed peak. 

It is interesting to note that in the context of the conductivity calculated from the band structure with canted AFM order, the origin of the peak at 200 cm$^{-1}$ is somewhat different than in the context of the uncanted structure, despite appearing at a similar energy scale. A decomposition into different inter-band contribution indicates that in this structure the peak is primarily from transitions between the previously degenerate bands, i.e., bands 37$\rightarrow$38 and 39$\rightarrow$40 (Band labels given in Fig. \ref{fig:bs}). Although there is not much evidence from our experiment to choose one over the other, we do not favor this as the correct interpretation of our data as a canted order is not supported by neutron scattering experiments or theoretical considerations. However, this calculation further elucidates the fact that this peak is not a signature of Weyl node, even in a band structure that incorporates the effects of canted AFM order.

A possible explanation for the discrepancy in amplitude of the 200 cm$^{-1}$ peak would be electronic correlation effects which can potentially add to the intra-band conductivity but would not be captured through simple subtraction of the Drude contribution. As an alternative to this peak being an inter-band excitation, it may arise from the coupling of conduction electrons to magnetism. Neutron scattering experiments on YbMnBi$_2$ \cite{FuhrmanPrivateCommunication} show prominent spin wave excitations in the energy range $\sim$ 120 - 340 cm$^{-1}$ which matches the energy scale of the low energy peak observed in conductivity. The extent of this coupling is hard to estimate solely from optical conductivity measurements. One can expect such correlations to redistribute the Drude spectral weight giving an effective mass of the quasiparticles, $m^*$, different from the band mass, $m_b$. Such mass renormalizations can be extracted using extended Drude model analysis but as argued before, it would be unreliable in the present context because of the low lying inter-band transitions. However we can roughly characterize the renormalization effect by spectral weight analysis. We make use of the conductivity sum rule
\begin{equation}
\dfrac{Z_0}{\pi^2}\int_0^{\omega_c} \sigma_1(\omega) d\omega = \omega_p^2 = \dfrac{4\pi n e^2}{m_b}
\end{equation}
where $Z_0$ is vacuum impedance, $n$ is the carrier density and $\omega_c$ is chosen appropriately to include the bandwidth of interest. If we interpret both the Drude term and peak at 200 cm$^{-1}$ both as inter-band contributions, then the ratio of the total spectral weight to the Drude piece gives a measure of the renormalized mass. A naive estimate of this electron-magnon coupling constant from this exercise, $\gamma = m^*/m_b -1 = 0.4$. This is however a rather crude estimate as we know that there is a finite contribution to the conductivity at these frequencies from inter-band transitions and hence these numbers are only a rough guide. It would be interesting to estimate coupling constants from the neutron scattering data. 

\begin{figure}[h!!]
\begin{center}
\includegraphics[width=0.48\textwidth]{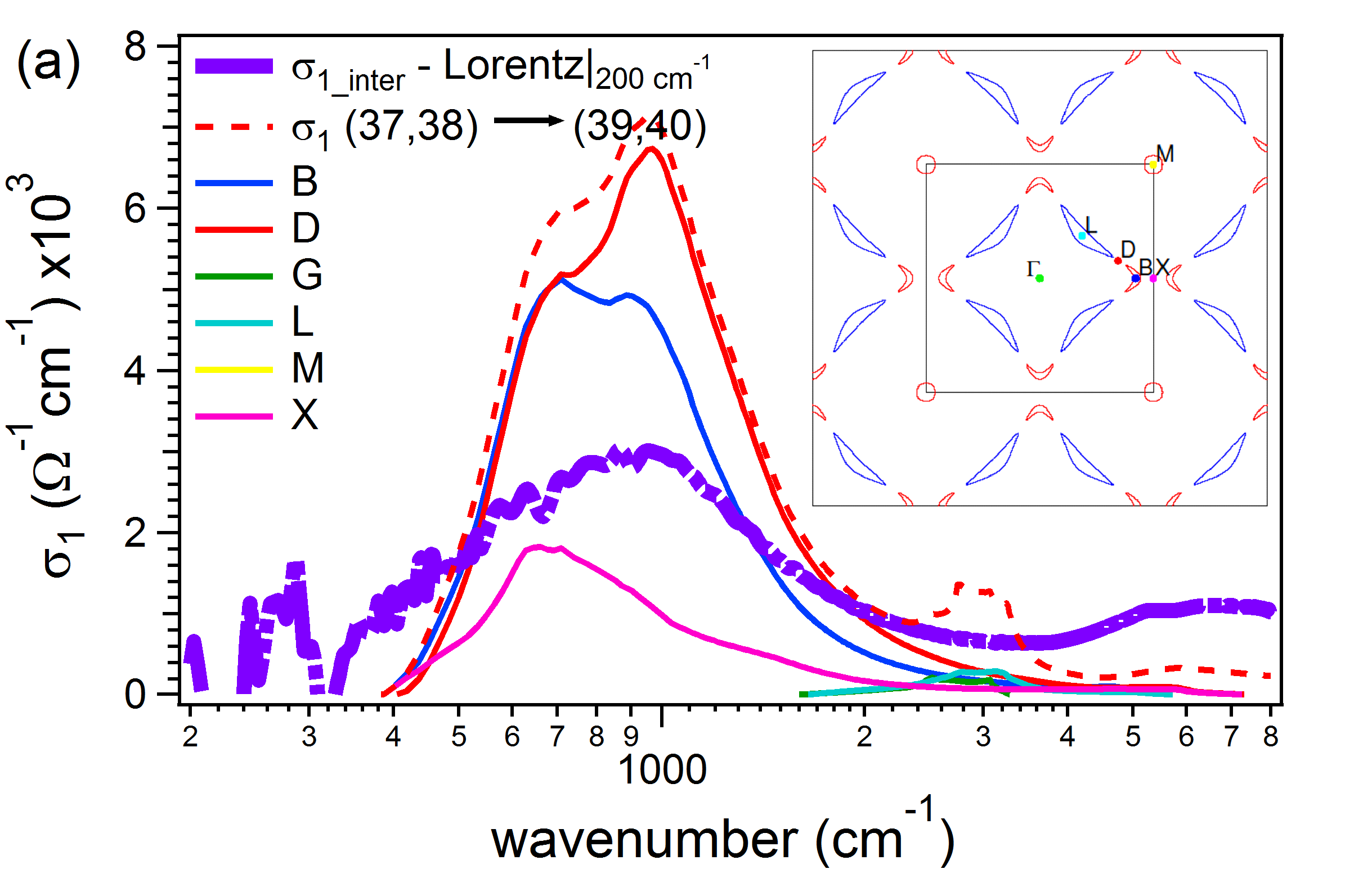}\\
\includegraphics[width=0.48\textwidth]{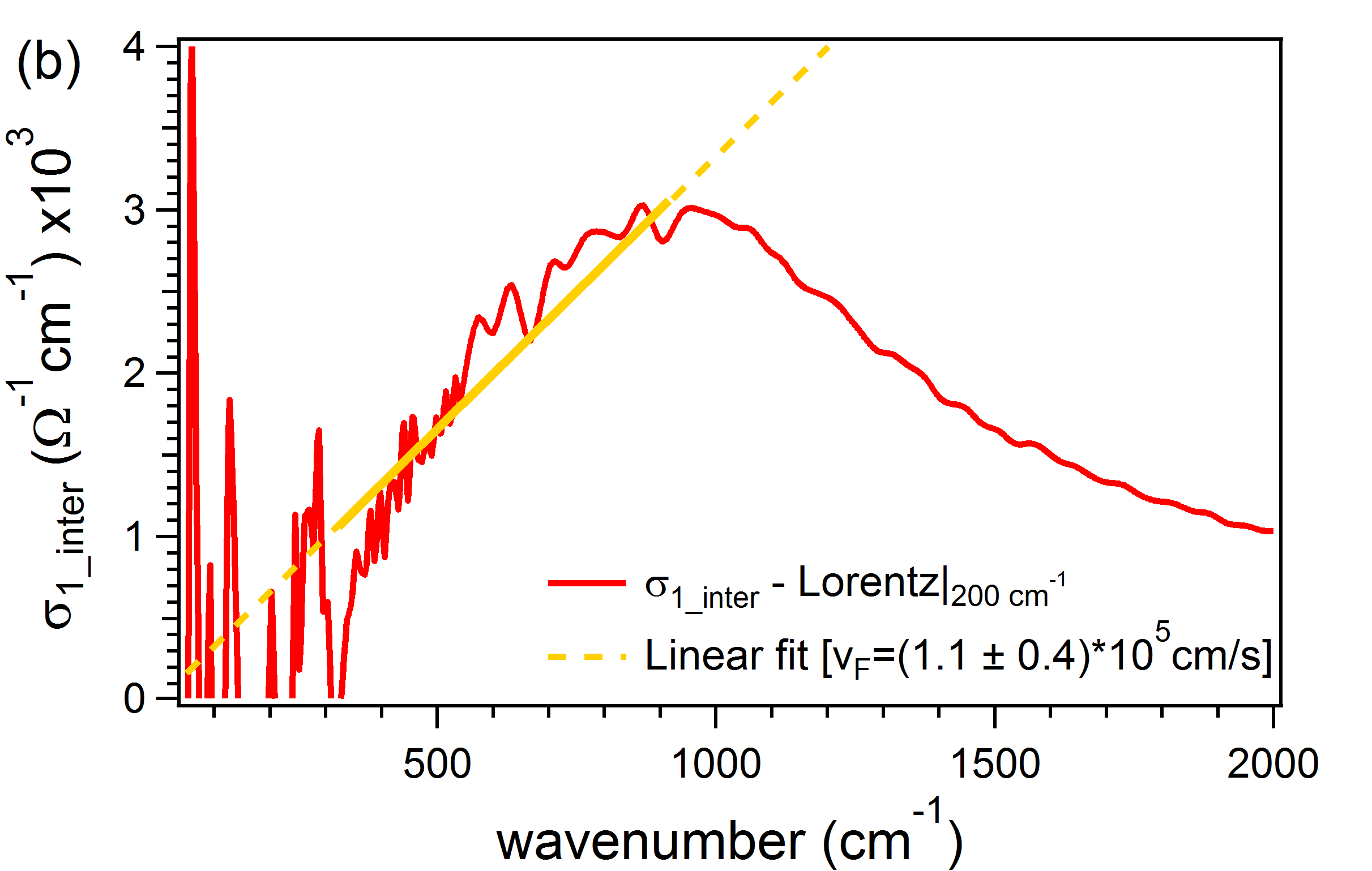}
\end{center}
\vspace{-10pt}
\caption{(a) Decomposition of the the inter-band contribution to the optical conductivity from the transition between the bands (37,38) $\rightarrow$ (39,40) from different parts of the Fermi surface (fig. inset) (b) Linear fit of the inter-band transition peaked at 950cm$^{-1}$.}\label{fig:dirac}
\vspace{-10pt}
\end{figure}

The peak at 950 cm$^{-1}$ is clearly from the inter-band transition between the bands of interest that would host the Weyl nodes in the canted structure i.e., right above (39,40) and below (37,38). However, because of the other low energy transition, it is nearly impossible to isolate any signature of the Weyl nodes if they exist in the optical conductivity spectrum. The peak derives from the quasi-2D Bi $p$-states (see Supplementary Materials) and is indicative of the linear dispersion. A sharp peak in conductivity is ordinarily expected to appear if a pair of occupied initial and empty final bands have nearly parallel dispersion in a large volume in $k$-space. Such an energy dispersion may occur if the Fermi level is inside a gap, which opens due to avoided crossing of two bands, and the size of the gap is constant in a large part of Brillouin zone. Hence it is reasonable to
conclude that this peak is a signature of a Dirac dispersion with a small mass gap. This gap between the ``lens'' shaped Fermi surfaces from the (37,38) bands and the ``boomerang'' shaped Fermi surfaces from the (39,40) bands (see figure \ref{fig:dirac}(a) (inset)) extends along $k_z$ over the whole Brillouin zone as the derived Bi $p$ bands are quasi-2D. This is further supported by partial conductivity calculations where we have integrated over cylindrical regions in $k$-space centered around points shown in the inset in Fig.\ \ref{fig:dirac}(a). The radius of each cylinder was chosen to be $\approx 0.1\times\frac{2\pi}{a}$ and the results clearly indicate that the major contribution to this peak is from the D-point which is between the ``lens'' and the ``boomerang'' shaped Fermi surfaces where the Fermi level lies in the gap. Calculations also indicate that the magnitude of this gap does not depend on $k_z$ as expected. The contribution from B and X-points are also quite significant but as these points are in close proximity to the D-point, it is not surprising because the cylindrical volumes are not mutually exclusive. This is the reason why these spectra are not additive. 

We can isolate the low frequency features of this particular inter-band contribution from the conductivity measured experimentally by fitting the peak at 200 cm$^{-1}$ to a Lorentz oscillator and subtracting it from the previously calculated inter-band conductivity. The resultant shows a roughly linear regime between 300 cm$^{-1}$ and 950 cm$^{-1}$ that can be fitted to a straight line with zero intercept (see figure \ref{fig:dirac}(b)). With the assumption that this linearity is due to the Dirac dispersion, we can calculate the Fermi velocity using equation \eqref{eq:lin}. There are four Dirac points in each Brillouin zone of YbMnBi$_2$ \cite{borisenko2015time} and thus the calculated Fermi velocity is (1.1$\pm$0.4)$\times$10$^5$ cm/s [or 0.045$\pm$0.01 eV$\cdot$\AA]. This estimated Fermi velocity is within the range expected from ARPES which varies between 9 eV$\cdot$\AA (perpendicular to the $k_xk_y$ plane) and 0.043 eV$\cdot$\AA (along $k_z$). As discussed before, anisotropic Dirac cones are a hallmark of AMnBi$_2$ compounds and our observations here on YbMnBi$_2$ fits well into this description. The existence of a Dirac fermionic dispersion is also in good agreement with the reported magneto-transport experiments \cite{PhysRevB.93.085442}.

We note that our interpretation of these peaks is at odds with another recent optical study \cite{chinotti2016electrodynamic} that has observed similar features in low energy conductivity but proposes that the peak around 200 cm$^{-1}$ derives from the van Hove singularity in a simplified WSM band structure. We do not favor this interpretation as realistic band structure calculations of YbMnBi$_2$ show it to exhibit certain non-idealities that will obscure the canonical behavior given in Ref. \cite{PhysRevB.93.085442}.  Moreover, the Weyl state appears to be dependent on the existence of the magnetic structure for which there is no evidence. To reiterate, the the gross features of the optical response can be explained by the electronic structure of a uncanted state with a slightly shifted chemical potential.

\section{Conclusion}
We have measured the reflectivity of YbMnBi$_2$ in the infrared regime and calculated the optical conductivity. The various features in the conductivity has been interpreted using DFT band structure calculations. We believe that the low energy peaks in conductivity at 200 cm$^{-1}$ and 950 cm$^{-1}$  are true inter-band transitions and the spectrum can be explained reasonably well without invoking a canted magnetic structure which is necessary to realize Weyl nodes in this system. The conductivity leading up to the peak at 950 cm$^{-1}$ is consistent with a linear Dirac dispersion with a small gap. Although canting of the Mn moments does improve the agreement of the calculated spectra with the experiments, similar effect can also be achieved by small shift in Fermi level. Precise control of the Fermi level in semimetals is rather difficult which when considering the lack evidence for canting, we believe makes the latter outcome more likely. However, in either case the high relative amplitude of the low energy peak is not apparent from band structure calculations but could be an manifestation of electronic correlation effects. 

One must still reconcile the ARPES data that shows reasonably convincing evidence for Weyl physics \cite{borisenko2015time}, with neutron scattering experiments \citep{FuhrmanPrivateCommunication, wang2016two} that does not show evidence for canting and our optical data that does not need it. Moreover, we should reiterate that the crystal structure of this compound does not allow for a canted magnetic structure to develop through a second order phase transition.  All of these considerations can be accommodated if we assume that the reduced symmetry of the surface allows a surface magnetic structure reconstruction that is consistent with the canted state and a surface Weyl phase. It is possible that the surface of this compound hosts a true WSM through the breaking of time-reversal symmetry. Detailed measurements of the surface electronic structure or magnetism would be very useful in this regard. 

\section{Acknowledgements}
We would like to thank C. Broholm, S. Borisenko and W. Fuhrman for helpful discussions and E. S. Arinze and N. Drichko for assistance with visible range reflectivity measurements. Optical measurements at JHU were supported by the U.S. Department of Energy, Office of Basic Energy Sciences under Contract DE-FG02-08ER46544.  Sample growth at Princeton was supported by the ARO MURI on topological insulators, Grant W911NF-12-0461.

\onecolumngrid
\newpage
\appendix
\setcounter{equation}{0}
\setcounter{figure}{0}
\setcounter{table}{0}
\setcounter{page}{1}
\makeatletter
\renewcommand{\theequation}{S\arabic{equation}}
\renewcommand{\thefigure}{S\arabic{figure}}
\renewcommand{\bibnumfmt}[1]{[S#1]}
\renewcommand{\citenumfont}[1]{S#1}
\section{\large{Supplimentary Materials: An optical investigation of the strong spin-orbital coupled magnetic semimetal YbMnBi$_2$}}

\section{Fermi surface map for the different band structures}
\begin{figure}[h!!]
\begin{center}
\includegraphics[width =0.35\textwidth]{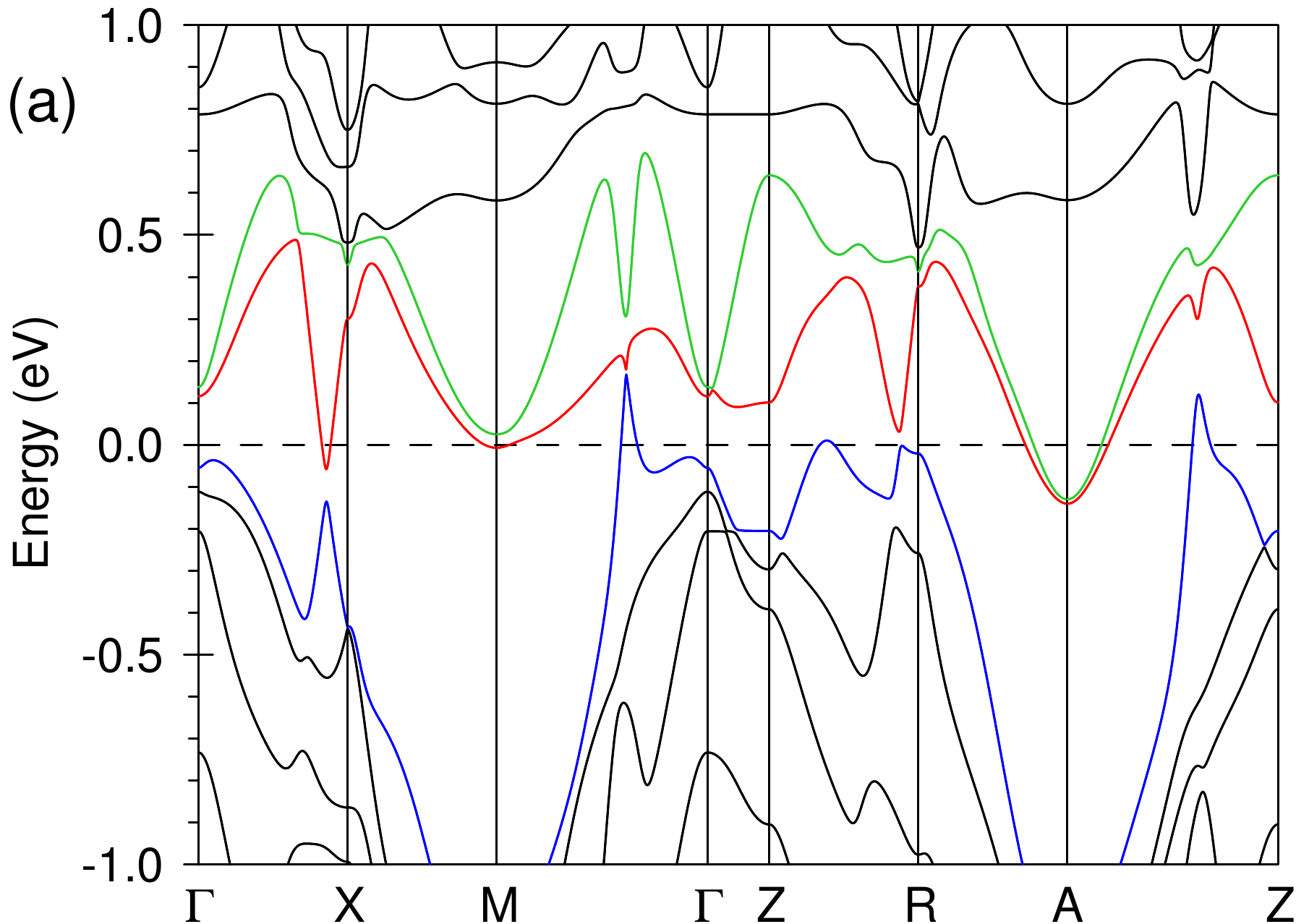}~~~~~~\includegraphics[width =0.25\textwidth]{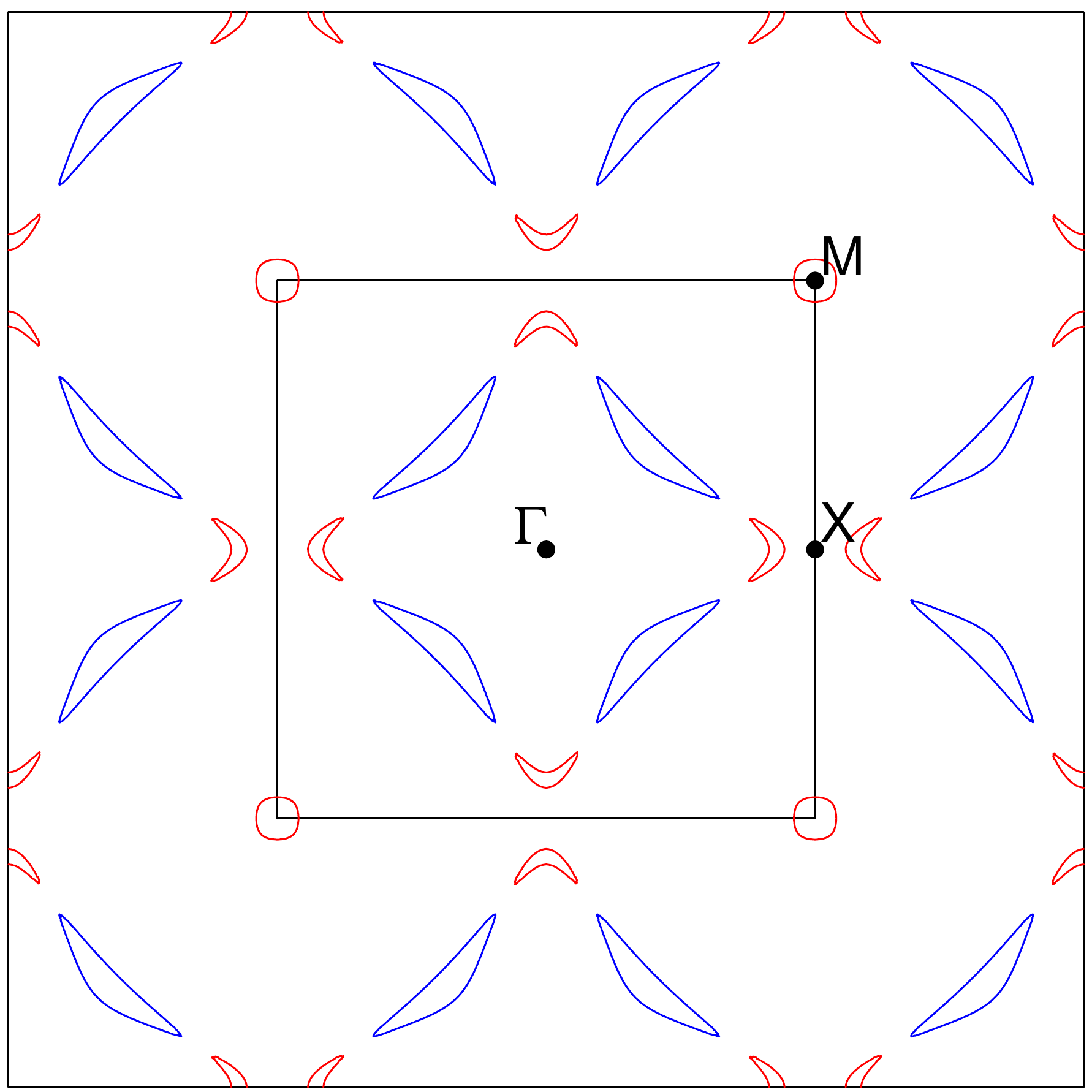}\\~\\
\includegraphics[width =0.35\textwidth]{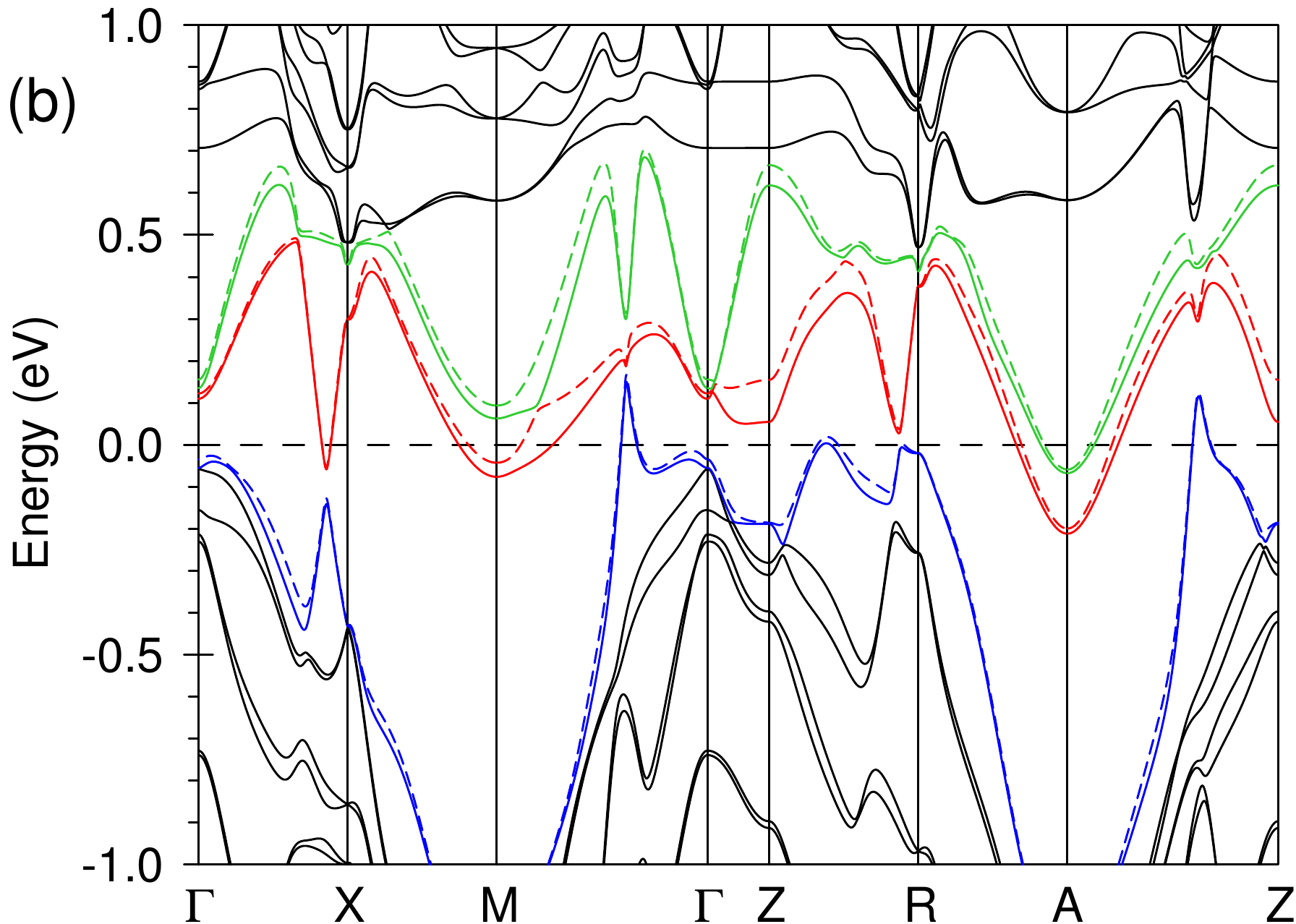}~~~~~~\includegraphics[width =0.25\textwidth]{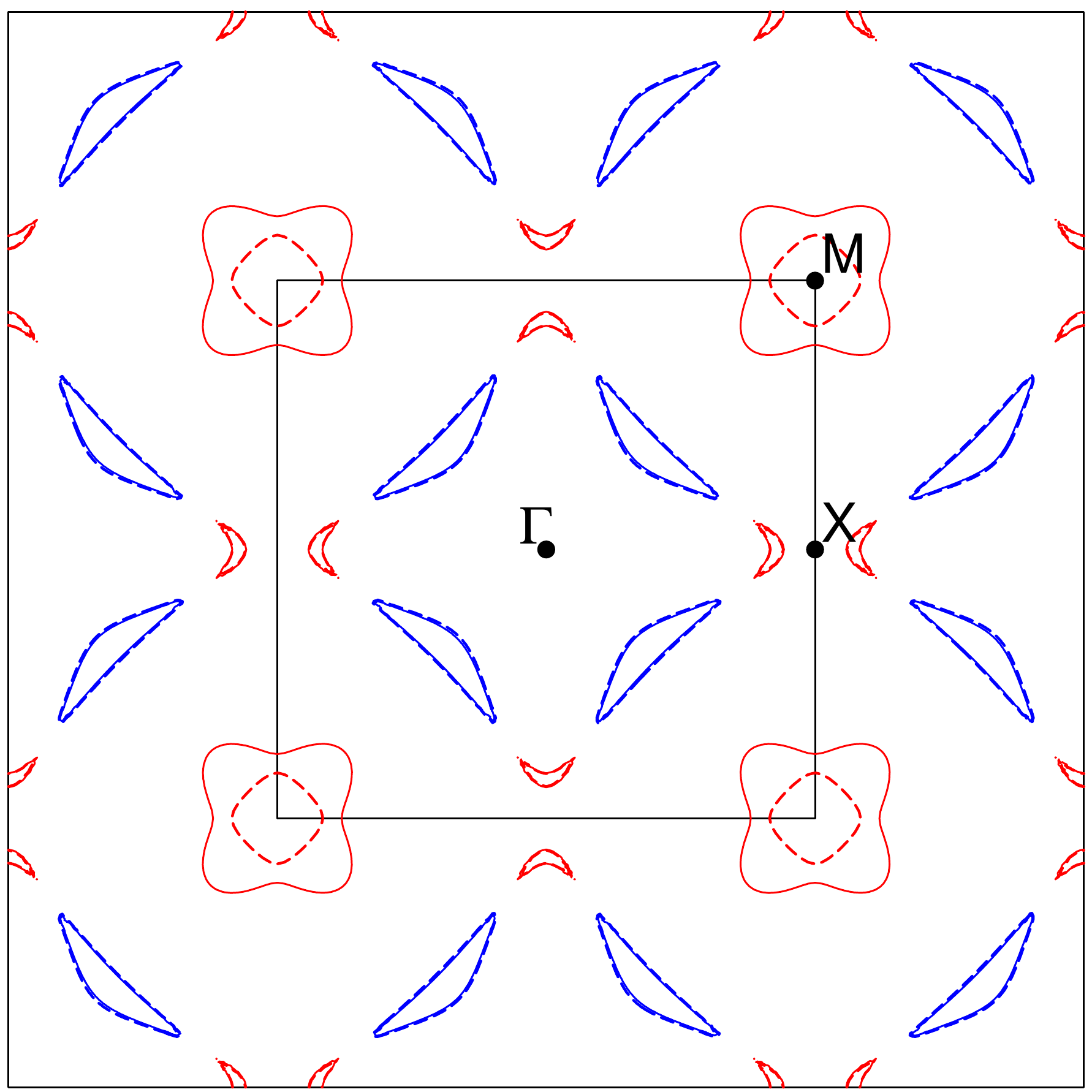}\\~\\
\includegraphics[width =0.35\textwidth]{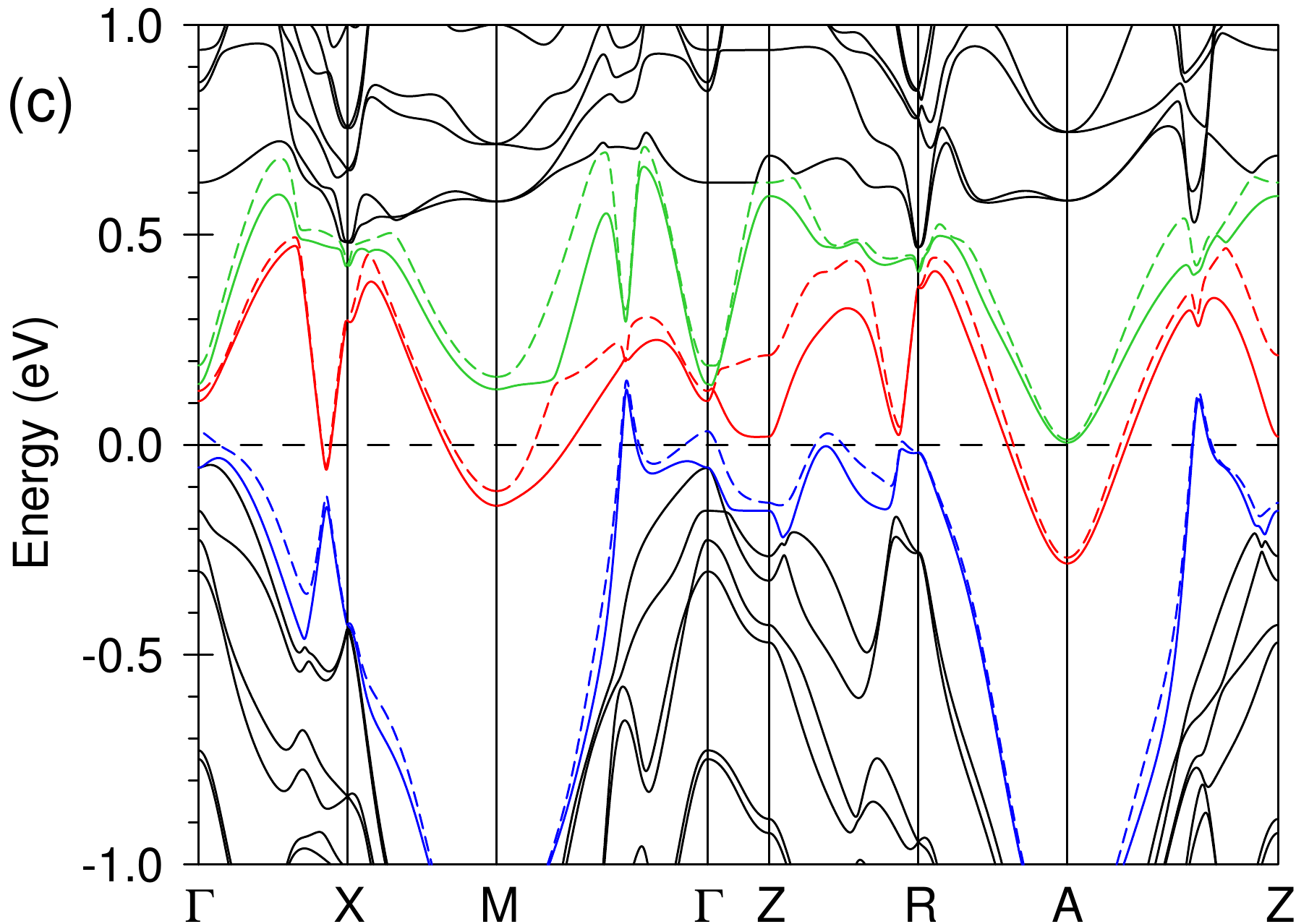}~~~~~~\includegraphics[width =0.25\textwidth]{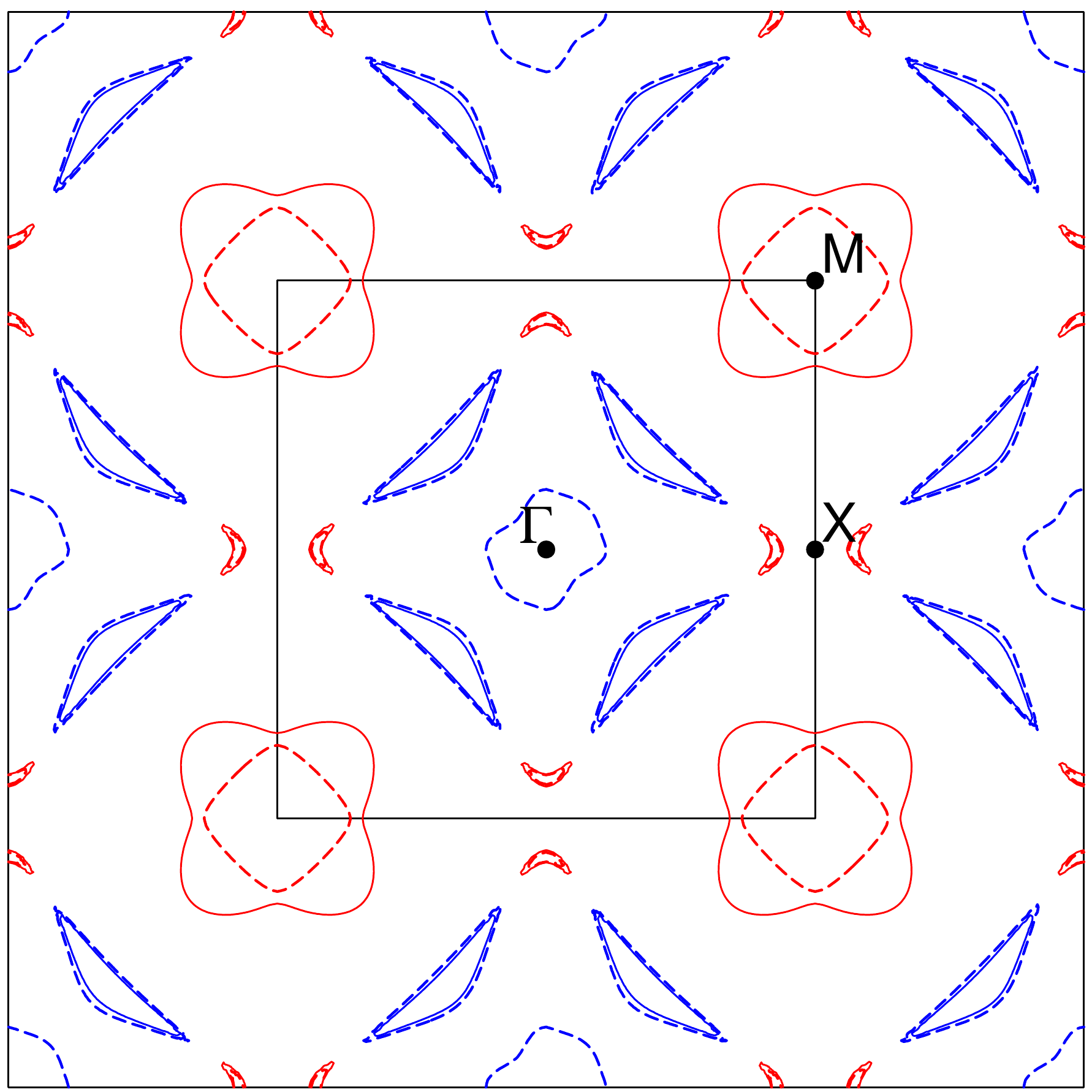}\\~\\
\end{center}
\vspace{-10pt}
\caption{Band structure along selected symmetry lines for a canting of (a) 0$^{\circ}$, (b) 5$^{\circ}$ and (c) 10$^{\circ}$ from the c-axis. The solid and the dashed lines in (b) and (c) of the same colour represent the bands that were degenerate in (a). Corresponding Fermi surface cross sections are plotted on the right.}\label{fig:fs}
\vspace{-10pt}
\end{figure}

The Fermi surface for $k_z=0$ corresponding to the different band structure calculations are plotted in fig. \ref{fig:fs}. 

\section{Fat bands}
\begin{figure}[h!!]
\begin{center}
\includegraphics[width=\textwidth]{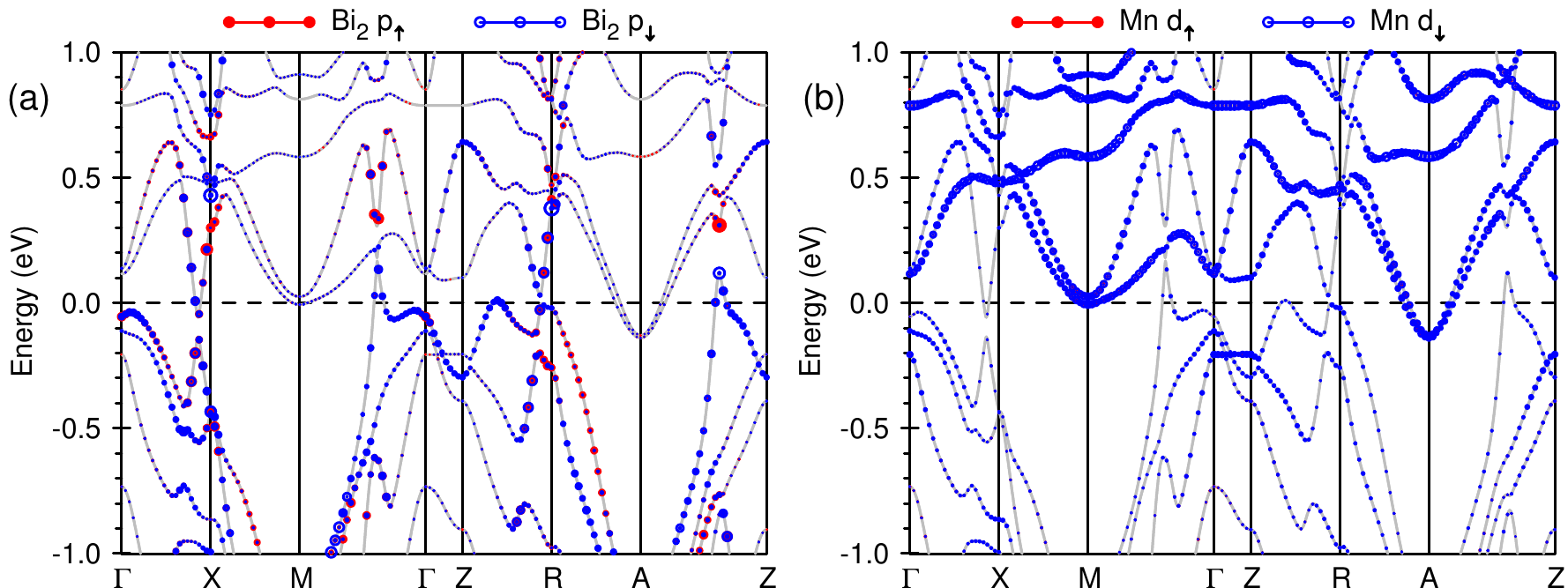}
\end{center}
\vspace{-10pt}
\caption{Band plots showing orbital contributions from (a) Bi $p$ states and (b) Mn $d$ states.}\label{fig:ocm}
\vspace{-10pt}
\end{figure}

~\\The origin of the peak in optical conductivity at 950 cm$^{-1}$, as discussed in the main text, is predominantly from the transition across the linearly dispersing bands. These massive Dirac like dispersion derives from the quasi-2D Bismuth $p$-states. So-called ``fat bands'' in Fig.\ \ref{fig:ocm} clearly show that the strongly dispersing bands are almost exclusively composed of Bi $p$ states [Fig.\ \ref{fig:ocm} (a)] whereas the relatively flat bands forming the electron pockets near the $Z$ point originate from Mn $d$ states [Fig.\ \ref{fig:ocm} (b)].\\~\\
It is worth mentioning, that the wave functions of the predominantly Bi $p$ derived bands have also contributions of Bi $s$ and $d$ states which allows dipole transitions between these bands.

\end{document}